# Characterizing the Variable X-Ray and UV–Optical Flux Behavior of Blazars

Kaitlyn E. Moo, Joel N. Bregman, and Mark T. Reynolds  
Department of Astronomy, University of Michigan, 1085 South University Avenue, Ann Arbor, MI 48109, USA

## Abstract

The variability of blazars in the X-ray and optical regions both informs the physics of their emitting region and places demands on the observer if a program requires that the object be bright or faint. The extensive simultaneous X-ray and optical observation by the Neil Gehrels Swift Observatory (Swift) provides the best insight into the variable nature of these objects. This program uses Swift data for 19 X-ray-bright blazars, generally at $z > 0.1$, to determine their variability properties. The analysis is based on structure functions and provides insight into the nature of the variability and how it depends on time, luminosity, and redshift. We also consider strategies for observing blazars at or above average brightness, given a time delay between planning an observation and obtaining the data. This is critical to observations with orbiting X-ray telescopes, current or future. The variability in the soft X-ray band is typically three to eight times larger than at UV–optical wavelengths, at fixed time differences (i.e., 30 or 100 days). There is almost no difference in the amplitude of variation (X-ray or UV–optical) as a function of redshift (time delay of 30 days) and a modest positive correlation with luminosity. In the X-ray band, blazars that become brighter than normal typically remain bright for at least 2–3 months, although with significant flickering. One can avoid observing objects that are significantly below the average X-ray flux by scheduling the observation when $F_X > 0.9 F_{X,avg}$, which requires monitoring observations near the time of the scheduling activity.

*Unified Astronomy Thesaurus concepts:* Blazars (164); X-ray active galactic nuclei (2035); X-ray telescopes (1825); Hot intergalactic medium (751)

*Supporting material:* figure sets

## 1. Introduction

The BL Lac objects and the optically violent variable classes of active galactic nuclei (AGNs), collectively referred to as blazars (Antonucci & Ulvestad 1985), are highly variable over the entire electromagnetic spectrum, with the greatest variation at the highest frequencies, such as the X-ray region (Pian 2002). This variability allows one to investigate the inner workings of the central engine, and the relationship between wave bands is obtained through multifrequency studies (Böttcher et al. 2003). They are often the brightest X-ray emitters at $z > 0.05$ and thus are excellent line-of-sight probes for studies of extragalactic absorption lines. Recent reviews of such objects, including discussions of emission processes, are found in Foschini (2017), Blandford et al. (2019), and Costamante (2020).

There are two purposes to this paper. One is the analysis of time variability properties in the UV–optical bands and X-ray bands for X-ray-bright blazars. This makes use of the extensive archive of blazars from Swift, which began observations in 2005, and builds on the effort of Stroh & Falcone (2013). The literature about blazar variability is large, with an emphasis on multiwavelength observations, and covers a wide range of timescales. A full review of the literature is not given here, but we try to give a sense of the efforts. In the past two decades, optical variations have been provided by worldwide networks of optical observatories, often using modest telescopes of ~1 m class, such as the Whole Earth Blazar Telescope program (Damljanović et al. 2018), the Tuorla Observatory blazar monitoring program (Katajainen et al. 1999), or individual dedicated telescopes, such as the Automatic Telescope for Optical Monitoring at the H.E.S.S. array (Wierzcholska et al. 2015). The launch of the Fermi Gamma-ray Space Telescope, a few years after Swift, made it easier to obtain observations in both the X-ray and gamma-ray regions. This led to a number of extensive investigations of variability from the radio through the gamma-ray region for individual objects, and on a broad range of timescales, such as for 3C-454.3, OJ 287, PKS 0735+178, PKS 2155-304, PKS 2004-489, Mrk 421, Mrk 501, and S5 0716+714, to mention a few (Raiteri et al. 2011; Aleksić et al. 2015; Ahnen et al. 2017; Goyal et al. 2017, 2018; Kapanadze et al. 2017; MAGIC Collaboration et al. 2018; Chevalier et al. 2019; Kapanadze 2021). For a sample of radio-bright blazars, we apply a uniform set of tests to characterize the variability, using structure functions to explore these properties.

The other purpose of the paper relates to the selection of targets for X-ray studies, and spectroscopic programs in particular. Observational programs for X-ray observatories are decided weeks or months in advance, due to the complicated nature of scheduling space observatories. For demanding and time-consuming programs, such as X-ray spectroscopy, one must avoid observations during low-flux periods. Here we examine whether the variable nature of blazars informs us as to how to avoid observations during low-flux periods. There are other types of observational programs in which one seeks to observe a blazar in its low state (e.g., Acciari et al. 2021), and this work can help to guide those studies as well.

Spectroscopic X-ray observations toward these blazars will be central in carrying out absorption line studies along cosmological lines of sight, providing insight into the properties of intervening hot gas ($10^{5.7}$–$10^7$ K). This line of study will complete the census of absorbing gas and the baryonic mass, complementing the work on cooler absorption line gas, mainly carried out in the UV and UV/optical bands (e.g., Shull et al. 2012; Tumlinson et al. 2017). Such X-ray spectroscopic observations have been performed with XMM-Newton and Chandra, which have detected hot Milky Way gas (e.g., Hodges-Kluck et al. 2016), but infrequently detect extragalactic absorption systems, and only at low signal-to-noise ratio (Nicastro et al. 2016). The detection of extragalactic X-ray absorption lines will improve greatly with the launch of Athena (Barcons et al. 2017) and Arcus (Smith et al. 2016) or Lynx (Gaskin et al. 2018), if approved. These offer more than an order-of-magnitude improvement in sensitivity, but the observing times will be lengthy, typically several days (100–1000 ks) per object (Bregman et al. 2015). The success of these studies depends on a targeted blazar not becoming particularly faint, which drives the question of characterizing the variability in the hope of avoiding a poor outcome.

One might monitor target blazars in advance of spectroscopic observations, but there will be a time gap between the last monitoring measurement and the X-ray spectroscopic observation, which may be weeks or months. We define the time between the last monitoring observation and the spectroscopic observation as $\Delta t$ and frame the question: is there a strategy to improve the likelihood of success of the X-ray spectroscopic observation?

Much of the emphasis of this work is on the X-ray variability as a guide for whether to schedule an observation. We also consider whether adopting a threshold value at the last monitoring observation is useful in deciding whether to schedule a spectroscopic observation. However, optical monitoring observations are easier to obtain than space-based X-ray measurements, so we also examine whether such optical data can be used to determine whether an X-ray spectroscopic observation should be scheduled. These issues can be addressed with time monitoring in both the X-ray and optical bands for the sources most likely to be observed for absorption line studies. The best targets are cataloged in Bregman et al. (2015), and there are excellent simultaneous X-ray and optical data for about 20 objects, obtained with the Neil Gehrels Swift Observatory (Swift) over the past 15 yr.

There has been some analysis of the variability of AGNs, but it is generally not useful for our needs. Much of the emphasis in the past two decades has focused on the time delay between different wave bands, often in nearby Seyfert galaxies (e.g., Breedt et al. 2010) but in only a few individual blazars, such as Mrk 421 (e.g., Zhang et al. 2019). The X-ray study of blazars has been greatly improved by Swift, where many of these objects have been observed multiple times (Stroh & Falcone 2013). There is an added advantage that the UV/optical instrument on Swift obtains simultaneous observations in the optical and/or UV bands.

Here we use the archival data accumulated by Swift, along with additional proprietary observations of the targets most likely to be observed early in the lifetime of missions such as Athena, Arcus, or Lynx. Following a statistical variability analysis of the X-ray data from the X-Ray Telescope (XRT) and the UV/optical data, we investigate strategies that would avoid observing targets when they are particularly faint. These strategies consider both X-ray monitoring and UV/optical monitoring options prior to making a decision to invest a significant amount of time for a spectroscopic observation.

**Table 1**
Swift Observatory UV–Optical Filters

| UV/Optical Filter | Central Wavelength (Å) | Width (Å) |
|---|---|---|
| W2 | 1928 | 657.0 |
| M2 | 2246 | 498.0 |
| W1 | 2600 | 693.0 |
| $U$ | 3465 | 785.0 |
| $B$ | 4392 | 975.0 |
| $V$ | 5468 | 769.0 |

**Note.** Each UV/optical filter on the Swift Observatory is listed (by ascending wavelength) with its central wavelength and full width at half maximum. The M2, W1, and W2 filters are considered to be in the near-UV wave band but are grouped with the $U$, $B$, and $V$ wave bands under the term UV/optical when compared to the X-ray data.

## 2. Source Selection and Data Processing

The sample of 19 AGNs are among the best sources against which to detect intergalactic absorption. These sources possess the highest product of flux and redshift, as listed in Bregman et al. (2015). The X-ray and UV/optical data were mostly obtained from the Swift Archive at the High Energy Astrophysics Science Archive Research Center (HEASARC),[1] along with some archived data from the Penn State University online database.[2] We reduced the data to factors that focused solely on time, UV/optical flux, X-ray flux, and error (positive and negative) pertaining to the aforementioned quantities.

The objects are mainly luminous BL Lacertae objects, with quasars that typically have observations for 2000+ days, and in a few cases observations for over 4500 days. They are observed through six UV/optical filters,[3] denoted by $U$, $V$, $B$, M2, W1, and W2, whose details are found in Table 1. For every X-ray observation, one also obtains a UV/optical observation in one or more of these bands. These UV/optical observations are analyzed separately from the X-rays, though they rise and fall as a group in terms of their flux variability. The bands may differ across the objects, as the "filter of the day" was chosen if a selection was not made by the observer (Troja 2020). As a result, some sources contain observations in all six UV/optical wavelengths, while others contain observations in only one or two of the bands.

The sources contain observations that range in date from 2005 March 31 to 2020 September 30, with the largest data set containing 361 individual observations; the lowest has 10. Table 2 provides information regarding the position, duration of observation, exposure time, wave band filter, and number of observations for each data set.

Of the 19 targets (excluding the stack), 17 are gamma-ray sources detected with the Large Area Telescope on the Fermi Gamma-ray Space Telescope, and sometimes with other gamma-ray observatories; the two sources that are undetected at gamma-ray energies are 1RXS J150759.8+041511 and 1RXS J15 (Stroh & Falcone 2013). The majority of sources (15/19) are classified as BL Lac objects, with the remaining four classified as flat-spectrum radio sources or QSOs (1RXS J150759.8+041511, 3C 273, 3C 454.3, and S5 0836+71).

---

[1] https://heasarc.gsfc.nasa.gov/cgi-bin/W3Browse/swift.pl
[2] https://www.swift.psu.edu/monitoring/
[3] https://www.mssl.ucl.ac.uk/www_astro/uvot/uvot_instrument/filterwheel/filterwheel.html##filterb

Table 2
AGN Object and Data Set Information

| Source Name | z | R.A. | Decl. | Start Date | Obs. Period | Filter | No. Obs. |
|---|---|---|---|---|---|---|---|
| 1RXS J003334.6-192130 | 0.61 | 00 33 34.38 | −19 21 33.14 | 11-09-2008 | 4024 | V | 19 |
| 1RXS J022716.6+020154 | 0.457 | 02 27 16.60 | +02 01 58.00 | 01-08-2018 | 2909 | B | 14 |
| S5 0836+71 | 2.17 | 08 41 24.37 | +70 53 42.17 | 04-13-2011 | 2001 | V | 51 |
| S5 0836+71 | 2.17 | 08 41 24.37 | +70 53 42.17 | 10-07-2011 | 1824 | M2 | 51 |
| S5 0836+71 | 2.17 | 08 41 24.37 | +70 53 42.17 | 04-13-2011 | 2001 | U | 53 |
| S5 0836+71 | 2.17 | 08 41 24.37 | +70 53 42.17 | 01-18-2009 | 2816 | W2 | 53 |
| S5 0836+71 | 2.17 | 08 41 24.37 | +70 53 42.17 | 01-23-2010 | 2446 | W1 | 54 |
| 1ES 1028+511 | 0.3604 | 10 31 18.54 | +50 53 35.82 | 05-28-2008 | 3645 | B | 14 |
| Mrk 421 | 0.031 | 11 04 27.31 | +38 12 31.80 | 04-22-2006 | 4798 | U | 16 |
| Mrk 421 | 0.031 | 11 04 27.31 | +38 12 31.80 | 03-31-2005 | 5134 | V | 17 |
| Mrk 421 | 0.031 | 11 04 27.31 | +38 12 31.80 | 03-31-2005 | 5199 | W1 | 767 |
| Mrk 421 | 0.031 | 11 04 27.31 | +38 12 31.80 | 03-31-2005 | 5199 | W2 | 813 |
| Mrk 421 | 0.031 | 11 04 27.31 | +38 12 31.80 | 03-31-2005 | 4800 | M2 | 931 |
| 1RXS J111706.3+201410 | 0.13793 | 11 17 06.25 | +20 14 07.38 | 04-20-2009 | 3883 | V | 15 |
| 1RXS J122121.7+301041 | 0.8136 | 12 21 21.94 | +30 10 37.16 | 10-30-2005 | 4552 | V | 88 |
| 1RXS J122121.7+301041 | 0.8136 | 12 21 21.94 | +30 10 37.16 | 10-30-2005 | 4552 | B | 90 |
| 1RXS J122121.7+301041 | 0.8136 | 12 21 21.94 | +30 10 37.16 | 10-30-2005 | 4814 | M2 | 91 |
| 1RXS J122121.7+301041 | 0.8136 | 12 21 21.94 | +30 10 37.16 | 10-30-2005 | 4552 | U | 92 |
| 1RXS J122121.7+301041 | 0.8136 | 12 21 21.94 | +30 10 37.16 | 10-30-2005 | 4959 | W1 | 93 |
| 1RXS J122121.7+301041 | 0.8136 | 12 21 21.94 | +30 10 37.16 | 10-30-2005 | 4957 | W2 | 96 |
| 3C 273 | 0.158339 | 12 29 06.70 | +02 03 07.60 | 02-23-2009 | 2727 | B | 14 |
| 3C 273 | 0.158339 | 12 29 06.70 | +02 03 07.60 | 02-02-2009 | 2719 | U | 71 |
| 3C 273 | 0.158339 | 12 29 06.70 | +02 03 07.60 | 07-10-2005 | 4370 | W1 | 75 |
| 3C 273 | 0.158339 | 12 29 06.70 | +02 03 07.60 | 07-10-2005 | 4051 | W2 | 86 |
| 3C 273 | 0.158339 | 12 29 06.70 | +02 03 07.60 | 07-10-2005 | 4370 | M2 | 91 |
| 3C 273 | 0.158339 | 12 29 06.70 | +02 03 07.60 | 07-10-2005 | 4051 | V | 100 |
| Ton 116 | 1.06625 | 12 43 12.74 | +36 27 44.00 | 02-13-2009 | 3379 | B | 13 |
| 2MASX J14283260+4240210 | 0.12926 | 14 28 32.61 | +42 40 21.05 | 03-31-2005 | 4771 | M2 | 145 |
| 2MASX J14283260+4240210 | 0.12926 | 14 28 32.61 | +42 40 21.05 | 03-31-2005 | 5302 | V | 148 |
| 2MASX J14283260+4240210 | 0.12926 | 14 28 32.61 | +42 40 21.05 | 03-07-2006 | 4430 | B | 153 |
| 2MASX J14283260+4240210 | 0.12926 | 14 28 32.61 | +42 40 21.05 | 03-31-2005 | 4771 | W1 | 155 |
| 2MASX J14283260+4240210 | 0.12926 | 14 28 32.61 | +42 40 21.05 | 03-31-2005 | 5302 | W2 | 156 |
| 2MASX J14283260+4240210 | 0.12926 | 14 28 32.61 | +42 40 21.05 | 03-31-2005 | 5302 | U | 157 |
| PG 1437+398 | 0.34366 | 14 39 17.48 | +39 32 42.81 | 10-15-2008 | 4237 | V | 22 |
| 1RXS J150759.8+041511 | 1.70057 | 15 07 59.73 | +04 15 11.98 | 08-05-2010 | 3336 | V | 15 |
| 1RXS J151747.3+652522 | 0.702 | 15 17 47.59 | +65 25 23.28 | 10-02-2014 | 979 | V | 34 |
| 1RXS J151747.3+652522 | 0.702 | 15 17 47.59 | +65 25 23.28 | 10-02-2014 | 979 | U | 35 |
| 1RXS J153501.1+532042 | 0.87496 | 15 35 00.80 | +53 20 37.31 | 06-19-2008 | 154 | B | 13 |
| PG 1553+113 | 0.36 | 15 55 43.04 | +11 11 24.37 | 03-05-2009 | 3867 | V | 248 |
| PG 1553+113 | 0.36 | 15 55 43.04 | +11 11 24.37 | 03-05-2009 | 3867 | B | 252 |
| PG 1553+113 | 0.36 | 15 55 43.04 | +11 11 24.37 | 03-05-2009 | 3867 | U | 253 |
| PG 1553+113 | 0.36 | 15 55 43.04 | +11 11 24.37 | 03-05-2009 | 3867 | M2 | 263 |
| PG 1553+113 | 0.36 | 15 55 43.04 | +11 11 24.37 | 03-05-2009 | 3867 | W1 | 268 |
| PG 1553+113 | 0.36 | 15 55 43.04 | +11 11 24.37 | 03-05-2009 | 3867 | W2 | 268 |
| PKS 2005-489 | 0.071 | 20 09 25.39 | −48 49 53.72 | 04-01-2011 | 2533 | V | 17 |
| PKS 2005-489 | 0.071 | 20 09 25.39 | −48 49 53.72 | 06-24-2009 | 3181 | U | 17 |
| PKS 2155-304 | 0.116 | 21 58 52.06 | −30 13 32.12 | 09-23-2009 | 2928 | U | 70 |
| PKS 2155-304 | 0.116 | 21 58 52.06 | −30 13 32.12 | 04-11-2006 | 4190 | V | 108 |
| 3C 454.3 | 0.859001 | 22 53 57.75 | +16 08 53.56 | 04-24-2005 | 4091 | V | 294 |
| 3C 454.3 | 0.859001 | 22 53 57.75 | +16 08 53.56 | 04-24-2005 | 4091 | B | 300 |
| 3C 454.3 | 0.859001 | 22 53 57.75 | +16 08 53.56 | 04-24-2005 | 4091 | U | 326 |
| 3C 454.3 | 0.859001 | 22 53 57.75 | +16 08 53.56 | 04-24-2005 | 4091 | M2 | 344 |
| 3C 454.3 | 0.859001 | 22 53 57.75 | +16 08 53.56 | 04-24-2005 | 4091 | W2 | 356 |
| 3C 454.3 | 0.859001 | 22 53 57.75 | +16 08 53.56 | 04-24-2005 | 4091 | W1 | 372 |
| H2356-309 | 0.16539 | 23 59 07.90 | −30 37 40.67 | 04-23-2012 | 3083 | V | 27 |
| Stack | ... | ... | ... | 05-28-2008 | 4024 | B, V | 243 |

**Note.** R.A. and decl. are given in J2000. Each data set used in our analysis is organized by source. The fifth column is the starting date of the observation, and the sixth column is the difference between the initial and final observing dates (in days). The typical observation time per object was 1700 days. Objects with more than 10 measurements were selected to obtain a high-quality UV/optical or X-ray observation for the variability, structure function, and triggering analyses. The UV/optical filter for the data set and the number of observations per data set are detailed in the seventh and eighth columns.

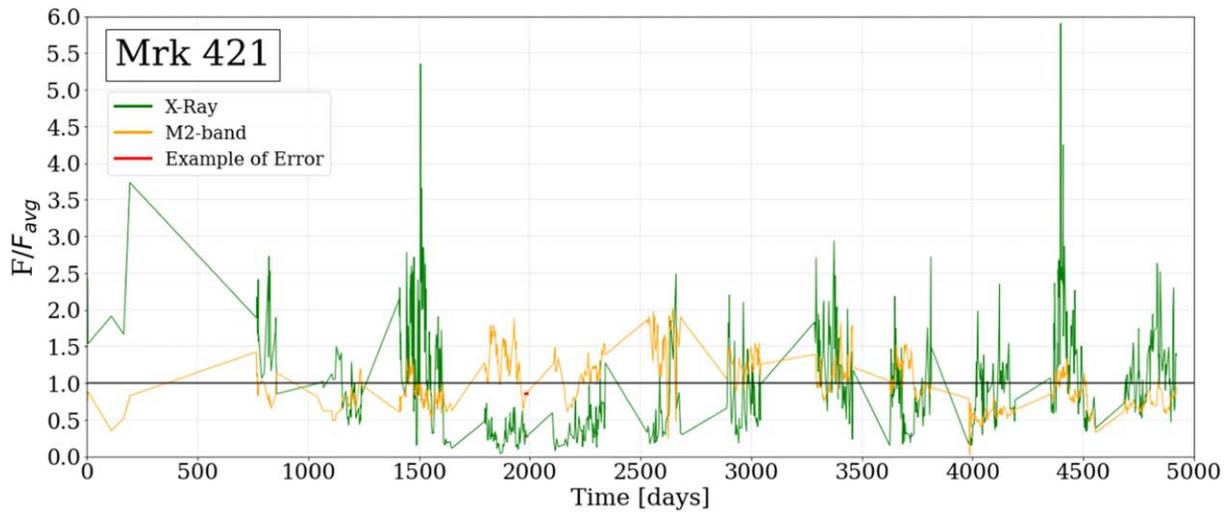
(a)

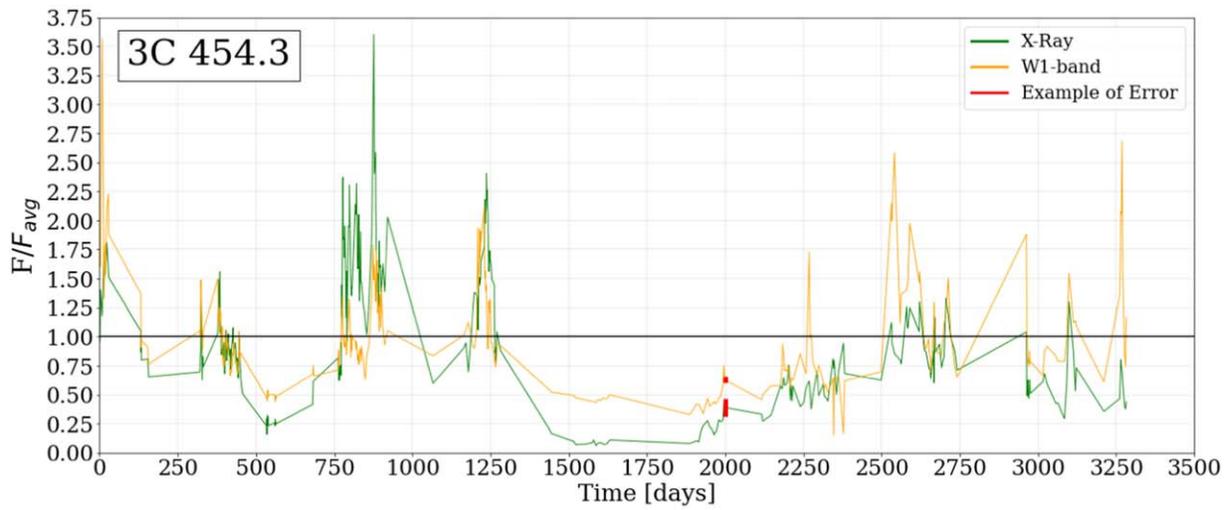
(b)

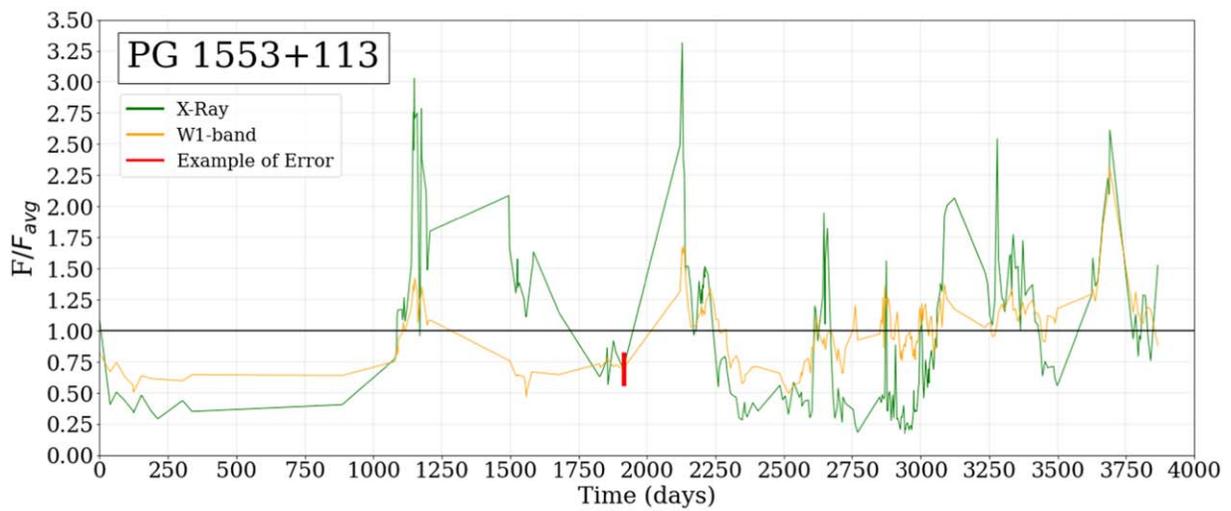
(c)

**Figure 1.** The three best documented objects, with typical error bars shown around 2000 days for each object. The black, horizontal line at 1.0 indicates a normalized flux value of 1.0.

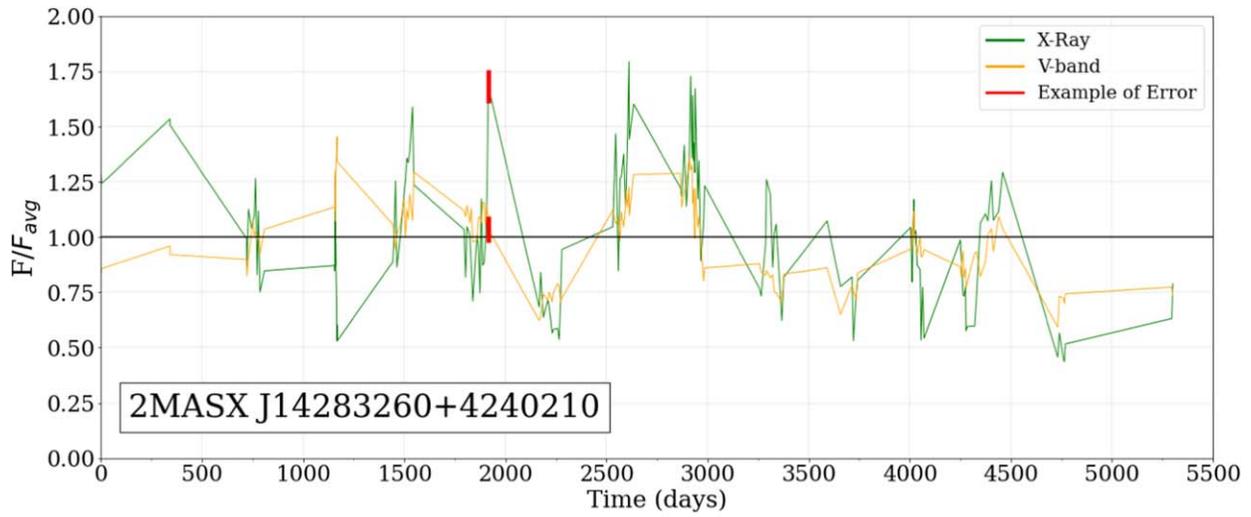

(d)

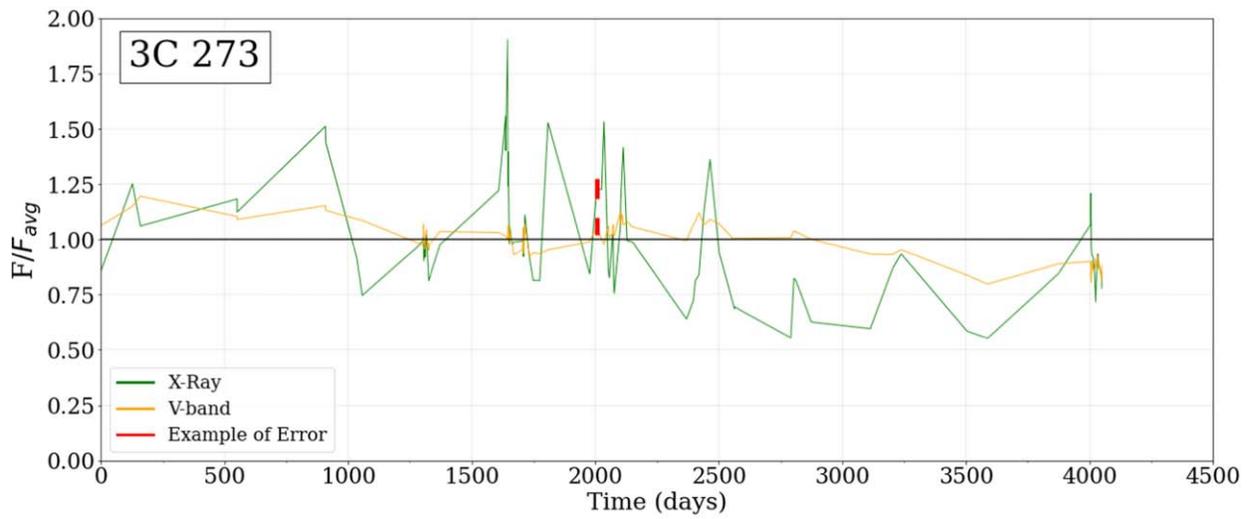

(e)

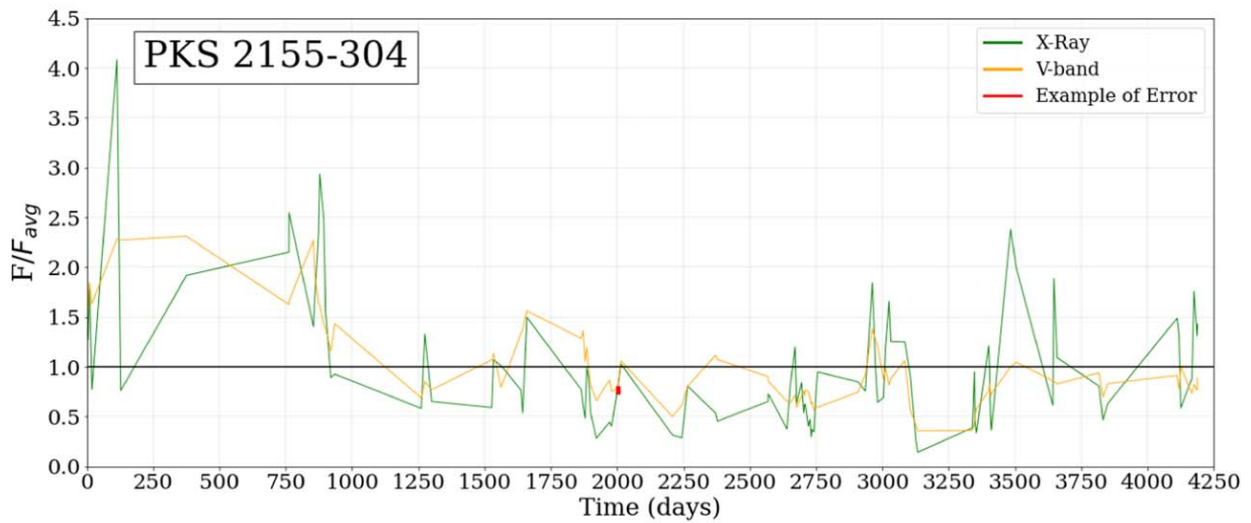

(f)

**Figure 2.** Fourth, fifth, and sixth best documented objects.

Table 3
AGN Object and Wave Band Flux Information

| Source Name | Filter | Mean Flux Density | | Median Flux Density | | $F_{\min}$ (Normalized) | | $F_{\max}$ (Normalized) | |
|---|---|---|---|---|---|---|---|---|---|
| | | X-ray | UV/Optical | X-ray | UV/Optical | X-ray | UV/Optical | X-ray | UV/Optical |
| 1RXS J003334.6-192130 | V | 5.10E-12 | 0.749 | 4.53E-12 | 0.622 | 0.412 | 0.641 | 2.360 | 1.571 |
| 1RXS J022716.6+020154 | B | 4.47E-12 | 0.137 | 3.56E-12 | 0.139 | 0.374 | 0.864 | 2.140 | 1.135 |
| S5 0836+71 | W2 | 2.70E-11 | 0.055 | 2.62E-11 | 0.055 | 0.513 | 0.818 | 1.552 | 1.254 |
| S5 0836+71 | M2 | 2.73E-11 | 0.077 | 2.68E-11 | 0.078 | 0.507 | 0.749 | 1.532 | 1.189 |
| S5 0836+71 | W1 | 2.68E-11 | 0.147 | 2.60E-11 | 0.145 | 0.512 | 0.698 | 1.547 | 1.246 |
| S5 0836+71 | U | 2.70E-11 | 0.456 | 2.62E-11 | 0.460 | 0.512 | 0.809 | 1.548 | 1.149 |
| S5 0836+71 | V | 2.72E-11 | 0.610 | 2.68E-11 | 0.614 | 0.509 | 0.775 | 1.539 | 1.174 |
| 1ES 1028+511 | B | 2.41E-11 | 0.473 | 2.40E-11 | 0.473 | 0.515 | 0.822 | 1.453 | 1.120 |
| Mrk 421 | W2 | 8.09E-10 | 9.975 | 7.13E-10 | 9.862 | 0.043 | 0.202 | 5.593 | 1.671 |
| Mrk 421 | M2 | 7.67E-10 | 11.980 | 6.60E-10 | 11.650 | 0.040 | 0.020 | 5.883 | 2.050 |
| Mrk 421 | W1 | 8.01E-10 | 11.400 | 7.15E-10 | 11.295 | 0.039 | 0.009 | 5.124 | 1.616 |
| Mrk 421 | U | 1.10E-09 | 8.400 | 1.24E-09 | 8.210 | 0.306 | 0.604 | 1.419 | 1.400 |
| Mrk 421 | V | 9.95E-10 | 17.860 | 8.11E-10 | 17.430 | 0.131 | 0.002 | 3.227 | 1.615 |
| 1RXS J111706.3+201410 | V | 4.12E-12 | 0.712 | 4.18E-12 | 0.670 | 0.178 | 0.866 | 2.006 | 1.209 |
| 1RXS J122121.7+301041 | W2 | 3.45E-11 | 0.384 | 3.41E-11 | 0.359 | 0.262 | 0.671 | 2.870 | 1.673 |
| 1RXS J122121.7+301041 | M2 | 3.41E-11 | 0.419 | 3.41E-11 | 0.419 | 0.264 | 0.538 | 2.892 | 1.656 |
| 1RXS J122121.7+301041 | W1 | 3.53E-11 | 0.463 | 3.42E-11 | 0.442 | 0.262 | 0.671 | 2.870 | 1.673 |
| 1RXS J122121.7+301041 | U | 3.53E-11 | 0.606 | 3.42E-11 | 0.579 | 0.264 | 0.640 | 2.894 | 1.613 |
| 1RXS J122121.7+301041 | B | 3.53E-11 | 0.739 | 3.41E-11 | 0.717 | 0.262 | 0.670 | 2.878 | 1.558 |
| 1RXS J122121.7+301041 | V | 3.43E-11 | 0.991 | 3.41E-11 | 0.977 | 0.262 | 0.630 | 2.883 | 1.481 |
| 3C 273 | W2 | 1.55E-10 | 11.746 | 1.53E-10 | 9.380 | 0.574 | 0.159 | 2.010 | 2.178 |
| 3C 273 | M2 | 1.56E-10 | 21.447 | 1.53E-10 | 21.210 | 0.583 | 0.768 | 2.006 | 1.234 |
| 3C 273 | W1 | 1.59E-10 | 22.407 | 1.58E-10 | 22.430 | 0.556 | 0.769 | 1.954 | 1.217 |
| 3C 273 | U | 1.66E-10 | 29.604 | 1.60E-10 | 29.425 | 0.544 | 0.814 | 2.870 | 1.161 |
| 3C 273 | B | 1.75E-10 | 26.423 | 1.60E-10 | 26.330 | 0.669 | 0.917 | 1.458 | 1.059 |
| 3C 273 | V | 1.63E-10 | 28.982 | 1.58E-10 | 29.170 | 0.553 | 0.798 | 1.903 | 1.195 |
| Ton 116 | B | 6.65E-12 | 0.995 | 6.41E-12 | 1.007 | 0.595 | 0.906 | 1.482 | 1.155 |
| 2MASX J14283260+4240210 | W2 | 4.94E-11 | 0.221 | 4.98E-11 | 0.224 | 0.435 | 0.592 | 1.788 | 1.491 |
| 2MASX J14283260+4240210 | M2 | 4.97E-11 | 0.243 | 4.99E-11 | 0.246 | 0.433 | 0.602 | 1.780 | 1.414 |
| 2MASX J14283260+4240210 | W1 | 4.95E-11 | 0.265 | 4.98E-11 | 0.274 | 0.435 | 0.573 | 1.786 | 1.439 |
| 2MASX J14283260+4240210 | U | 4.93E-11 | 0.359 | 4.96E-11 | 0.363 | 0.437 | 0.593 | 1.793 | 1.453 |
| 2MASX J14283260+4240210 | B | 4.93E-11 | 0.523 | 4.98E-11 | 0.533 | 0.436 | 0.664 | 1.792 | 1.283 |
| 2MASX J14283260+4240210 | V | 4.95E-11 | 0.911 | 4.98E-11 | 0.916 | 0.434 | 0.654 | 1.784 | 1.283 |
| PG 1437+398 | V | 3.36E-12 | 0.717 | 3.77E-12 | 0.718 | 0.346 | 0.857 | 1.776 | 1.138 |
| 1RXS J150759.8+041511 | V | 5.07E-12 | 2.016 | 5.52E-12 | 1.861 | 0.355 | 0.705 | 1.356 | 1.341 |
| 1RXS J151747.3+652522 | U | 1.16E-11 | 0.314 | 9.83E-12 | 0.315 | 0.399 | 0.809 | 2.734 | 1.270 |
| 1RXS J151747.3+652522 | V | 1.21E-11 | 0.440 | 1.02E-11 | 0.439 | 0.393 | 0.793 | 2.613 | 1.508 |
| 1RXS J153501.1+532042 | B | 9.51E-12 | 0.170 | 8.74E-12 | 0.175 | 0.667 | 0.841 | 1.427 | 1.075 |
| PG 1553+113 | W2 | 3.82E-11 | 2.511 | 3.53E-11 | 2.478 | 0.172 | 0.370 | 3.297 | 2.478 |
| PG 1553+113 | M2 | 3.80E-11 | 2.859 | 3.57E-11 | 2.871 | 0.171 | 0.346 | 3.285 | 2.266 |
| PG 1553+113 | W1 | 3.80E-11 | 3.426 | 3.51E-11 | 3.440 | 0.173 | 0.468 | 3.316 | 2.307 |
| PG 1553+113 | U | 3.59E-11 | 4.783 | 3.28E-11 | 4.763 | 0.182 | 0.461 | 3.478 | 1.860 |
| PG 1553+113 | B | 3.63E-11 | 5.932 | 3.32E-11 | 5.956 | 0.180 | 0.432 | 3.457 | 1.999 |
| PG 1553+113 | V | 3.65E-11 | 7.321 | 3.36E-11 | 7.387 | 0.180 | 0.550 | 3.457 | 1.772 |
| PKS 2005-489 | U | 1.99E-11 | 4.167 | 1.88E-11 | 3.383 | 0.628 | 0.654 | 1.608 | 1.473 |
| PKS 2005-489 | V | 2.02E-11 | 6.859 | 1.92E-11 | 5.638 | 0.619 | 0.747 | 1.584 | 1.522 |
| PKS 2155-304 | U | 4.11E-11 | 7.609 | 3.48E-11 | 7.589 | 0.186 | 0.469 | 2.443 | 1.440 |
| PKS 2155-304 | V | 5.33E-11 | 14.146 | 4.30E-11 | 12.100 | 0.140 | 0.353 | 3.991 | 2.536 |
| 3C 454.3 | W2 | 5.05E-11 | 0.477 | 4.46E-11 | 0.438 | 0.064 | 0.252 | 3.849 | 3.857 |
| 3C 454.3 | M2 | 5.16E-11 | 0.607 | 4.61E-11 | 0.560 | 0.063 | 0.200 | 3.681 | 3.467 |
| 3C 454.3 | W1 | 5.20E-11 | 0.820 | 4.59E-11 | 0.743 | 0.062 | 0.148 | 3.765 | 3.547 |
| 3C 454.3 | U | 4.58E-11 | 1.389 | 4.19E-11 | 1.255 | 0.070 | 0.086 | 4.126 | 4.506 |
| 3C 454.3 | V | 4.55E-11 | 3.024 | 4.20E-11 | 2.869 | 0.072 | 0.040 | 3.446 | 3.323 |
| 3C 454.3 | B | 4.56E-11 | 2.009 | 4.20E-11 | 1.889 | 0.072 | 0.061 | 3.447 | 3.197 |
| H2356-309 | V | 3.00E-11 | 0.552 | 2.60E-11 | 0.565 | 0.395 | 0.810 | 1.948 | 1.184 |
| Stack | B, V | ⋯ | ⋯ | ⋯ | ⋯ | 0.179 | 0.641 | 2.920 | 1.815 |

**Note.** All objects with mean, median, minimum, and maximum flux values for each data set (filter), excluding outliers. The UV or optical flux densities are in mJy while the X-ray flux densities are given in erg cm$^{-2}$ s$^{-1}$. The energy band in which the X-ray flux has been estimated is the standard energy range for the soft X-ray telescope (0.5–10 keV). The minimum and maximum fluxes are given in units normalized to the mean of each respective data set.

**Table 4**
60 days SF Linear Curve Fits: Top Six Sources

| Object | Wave Band | $k_1$ | ± | $a_1$ | ± |
|---|---|---|---|---|---|
| Mrk 421 | M2 | 1.24E-03 | 9.48E-05 | 3.33E-02 | 3.08E-03 |
| Mrk 421 | X-ray | 6.50E-03 | 1.51E-03 | 5.00E-01 | 4.89E-02 |
| 3C 454.3 | W1 | 3.77E-03 | 3.04E-04 | 6.40E-02 | 9.87E-03 |
| 3C 454.3 | X-ray | 3.56E-03 | 7.58E-04 | 2.06E-01 | 2.46E-02 |
| PG 1553+113 | W1 | 1.05E-03 | 8.67E-05 | 9.93E-03 | 2.82E-03 |
| PG 1553+113 | X-ray | 7.66E-03 | 7.71E-04 | 7.12E-02 | 2.50E-02 |
| 2MASX J14283260 +4240210 | U | 2.05E-04 | 7.34E-05 | 1.19E-02 | 2.39E-03 |
| 2MASX J14283260 +4240210 | X-ray | 1.10E-03 | 2.39E-04 | 4.66E-02 | 7.75E-03 |
| 3C 273 | V | 1.89E-05 | 1.34E-05 | 2.59E-03 | 4.35E-04 |
| 3C 273 | X-ray | 6.09E-04 | 4.36E-04 | 5.31E-02 | 1.42E-02 |
| PKS 2155-304 | V | 1.96E-04 | 5.45E-04 | 3.55E-03 | 1.80E-02 |
| PKS 2155-304 | X-ray | 3.17E-03 | 2.35E-03 | 2.33E-01 | 7.78E-02 |

**Note.** Unitless curve fit parameters for Figure 3.

## 3. Data Reduction

Final archived versions of the observation event and related calibration files were downloaded from the HEASARC[4] database and reprocessed with the latest version of the appropriate calibration files. Specifically, the XRT observations were reprocessed with XRTPIPELINE. Spectra were extracted using appropriate regions for observations in windowed timing (WT) and photon counting (PC) mode respectively,[5] where background spectra were extracted from neighboring source-free regions of the detector. Ancillary response files were generated with XRTMKARF, while the appropriate response file from the Swift calibration database was utilized. In observations where pile-up is an issue (Davis 2001), standard mitigation strategies are adopted. Spectra are instead extracted from annular extraction regions, where progressively larger regions of the piled-up core of the point-spread function are excised as count rates increase (see Reynolds & Miller 2013 for further details). In PC mode pile-up occurs when sources are brighter than $\gtrsim 1$ counts s$^{-1}$, whereas in WT mode pile-up becomes problematic for count rates $\gtrsim 100$ counts s$^{-1}$. Each observation was imported and analyzed in XSPEC V12.10.1F (Arnaud 1996).

Spectra were grouped to 20 counts bin$^{-1}$ using GRPPHA for spectra with more than 500 counts and the $\chi^2$ statistic was used. Below this threshold, spectra were grouped to 1 count bin$^{-1}$ and the C-statistic was used. The spectra were characterized with a simple power-law model (pha(po)), where the line-of-sight absorption was constrained to be larger than the Galactic value obtained from the HI4PI Map (HI4PI Collaboration et al. 2016)[6] and the power-law index was constrained to be $1 \leqslant \Gamma \leqslant 5$. All spectral fits were carried out in the 0.5–10.0 keV energy range. Parameter uncertainties were estimated with the ERROR command (errors are quoted at the 90% confidence level (1.645σ) unless otherwise noted) and observed fluxes were calculated with the FLUX command.

---

[4] https://heasarc.gsfc.nasa.gov
[5] https://www.swift.ac.uk/analysis/xrt/
[6] https://heasarc.gsfc.nasa.gov/cgi-bin/Tools/w3nh/w3nh.pl

Photometry was carried out on the simultaneous UV/optical imaging observations using the UVOTMAGHIST task. The filters utilized are described in Table 1 and a detailed breakdown of the filters utilized for each source is presented in Table 2. Source flux was extracted from a region of 5″ radius centered on the known source coordinates, with background estimated from a neighboring source-free region. The UVOT fields were inspected to ensure the absence of sources in the background regions for each observation.

### 3.1. Normalization Process

We normalized our data with respect to the average flux for each data set. Outliers in both UV/optical and X-ray flux were eliminated from our data sets for all parts of our analysis utilizing the interquartile range (IQR). Such outliers were calculated to be observations that are either less than the difference between the median and 1.5× the IQR or greater than the sum of the median and 1.5× the IQR (Dovoedo & Chakraborti 2015):

$$\text{lower outlier} = Q_1 - (1.5 \times \text{IQR}) \quad (1)$$

$$\text{upper outlier} = Q_3 + (1.5 \times \text{IQR}). \quad (2)$$

The outliers tended to be very distant from most of the observations in the data sets, and changing the factor of the IQR to values such as 1.4 and 1.6 yielded no difference in the analysis of the resulting data set. These points were eliminated if not found within a cluster of similar flux densities (20% above and below the flux of the observation in question within 10 days) that would indicate an X-ray flare or a duration of unusually high or low emission.

### 3.2. Time-delay Method

We utilized a time-delay method for each data set to gain a better understanding of flux changes on timescales of varying length. This was accomplished by calculating the time differences between one observation and all other observations within a data set of length $n$, given by

$$\Delta t = t_n - t_{n-1}. \quad (3)$$

The result from one iteration is an array of time offsets that contains the differences between point $n$ in the data set and all other observations in the set. We repeated this process for each data point until a set of arrays containing the time differences between all observations was created, along with their corresponding fluxes. We sorted this array from the smallest to the greatest time difference to gauge the behavior of the source over $\Delta t$ days.

### 3.3. Stacking

Data sets with less than 50 observations were "stacked" into a group, in which the UV/optical and X-ray flux densities have been normalized by their respective average fluxes. Flux differences were determined as a function of the time difference ($\Delta t$) for each object, then combined afterwards. The stacking of these data sets results in a source containing 243 data points that is similar to a typical data set, allowing us to have sufficient sampling to compare sources with smaller data sets. The sources and their respective filters/bands are listed as: 1ES 1028 (*B*), 1RXS J003334.6-192130 (*V*), 1RXS J022716.6+020154 (*B*), 1RXS J111706.3+201410 (*V*), 1RXS J150759.8+041511 (*V*),

Table 5
SF Curve Fit Parameters and Uncertainties

| Source Name | Wave band | $\Delta t < 1$ day | $\Delta t$ | xi | ± | $k_1$ | ± | $k_2$ | ± | $a_1$ | ± | SF (30 $\Delta t$) | $\sqrt{SF}$ (30 $\Delta t$) | SF (100 $\Delta t$) | $\sqrt{SF}$ (100 $\Delta t$) |
|---|---|---|---|---|---|---|---|---|---|---|---|---|---|---|---|
| 1RXS J003334.6-192130 | V | N | 175 | ⋯ | ⋯ | 0.014 | 0.196 | ⋯ | ⋯ | −2.527 | 0.347 | 0.005 | 0.071 | 0.006 | 0.078 |
| 1RXS J003334.6-192130 | X-ray | N | 175 | 0.845 | 0.345 | 1.835 | 1.815 | 0.000 | 0.287 | −2.179 | 1.154 | 0.242 | 0.492 | 0.242 | 0.492 |
| 1RXS J022716.6+020154 | B | N | 205 | ⋯ | ⋯ | 0.274 | 0.293 | ⋯ | ⋯ | −2.658 | 0.563 | 0.004 | 0.059 | 0.006 | 0.075 |
| 1RXS J022716.6+020154 | X-ray | N | 205 | ⋯ | ⋯ | 0.000 | 0.291 | ⋯ | ⋯ | −0.655 | 0.556 | 0.159 | 0.399 | 0.159 | 0.399 |
| S5 0836+71 | W2 | N | 200 | 1.898 | 0.946 | 0.273 | 0.233 | 0.041 | 0.701 | −2.526 | 0.361 | 0.006 | 0.078 | 0.008 | 0.089 |
| S5 0836+71 | X-ray | N | 200 | 1.913 | 0.270 | 0.743 | 0.206 | 0.010 | 0.672 | −2.373 | 0.323 | 0.034 | 0.185 | 0.066 | 0.257 |
| 1ES 1028 | B | N | 236 | 1.929 | 1.488 | 0.615 | 0.445 | 0.341 | 2.184 | −3.113 | 0.733 | 0.006 | 0.078 | 0.013 | 0.112 |
| 1ES 1028 | X-ray | N | 236 | 2.053 | 0.645 | 0.856 | 0.412 | 0.228 | 4.090 | −2.213 | 0.702 | 0.112 | 0.334 | 0.261 | 0.511 |
| Mrk 421 | M2 | Y | 500 | 0.784 | 0.173 | 0.664 | 0.050 | 0.391 | 0.010 | −1.92 | 0.030 | 0.076 | 0.276 | 0.114 | 0.337 |
| Mrk 421 | X-ray | Y | 500 | 0.305 | 0.097 | 1.037 | 0.110 | 0.149 | 0.012 | −0.663 | 0.059 | 0.705 | 0.839 | 0.816 | 0.903 |
| 1RXS J111706.3+201410 | V | N | 167 | ⋯ | ⋯ | 1.041 | 0.181 | ⋯ | ⋯ | −3.982 | 0.324 | 0.007 | 0.082 | 0.008 | 0.091 |
| 1RXS J111706.3+201410 | X-ray | N | 167 | ⋯ | ⋯ | 0.144 | 0.204 | ⋯ | ⋯ | −1.096 | 0.364 | 0.115 | 0.339 | 0.158 | 0.398 |
| 1RXS J122121.7+301041 | W2 | N | 200 | 1.452 | 0.832 | 0.541 | 0.296 | 0.299 | 0.233 | −2.50 | 0.330 | 0.018 | 0.134 | 0.025 | 0.159 |
| 1RXS J122121.7+301041 | X-ray | N | 200 | 1.000 | 0.650 | 0.454 | 0.582 | 0.000 | 0.152 | −1.179 | 0.413 | 0.185 | 0.430 | 0.204 | 0.452 |
| 3C 273 | V | Y | 200 | 0.000 | 1.994 | 0.302 | 0.835 | 0.032 | 0.126 | −2.631 | 0.488 | 0.003 | 0.051 | 0.003 | 0.052 |
| 3C 273 | X-ray | Y | 200 | 0.699 | 0.709 | 0.734 | 0.516 | 0.000 | 0.172 | −1.736 | 0.279 | 0.056 | 0.237 | 0.056 | 0.237 |
| Ton 116 | B | N | 154 | ⋯ | ⋯ | 1.29 | 0.376 | ⋯ | ⋯ | −4.387 | 0.676 | 0.003 | 0.055 | 0.018 | 0.135 |
| Ton 116 | X-ray | N | 154 | ⋯ | ⋯ | 0.968 | 0.334 | ⋯ | ⋯ | −2.797 | 0.601 | 0.041 | 0.202 | 0.151 | 0.389 |
| 2MASX J14283260+4240210 | U | Y | 150 | 0.661 | 0.621 | 0.635 | 0.339 | 0.0911 | 0.128 | −2.295 | 0.183 | 0.016 | 0.126 | 0.018 | 0.133 |
| 2MASX J14283260+4240210 | X-ray | Y | 150 | 0.716 | 0.270 | 0.925 | 0.214 | 0.131 | 0.094 | −1.870 | 0.120 | 0.078 | 0.279 | 0.091 | 0.302 |
| PG 1437+389 | V | N | 250 | 1.204 | 1.397 | 1.249 | 11.823 | 0.045 | 0.189 | −3.808 | 8.745 | 0.003 | 0.055 | 0.006 | 0.078 |
| PG 1437+389 | X-ray | N | 250 | 1.041 | 2.818 | 0.957 | 3.994 | 0.606 | 0.277 | −2.119 | 3.346 | 0.138 | 0.372 | 0.289 | 0.538 |
| 1RXS J150759.8+041511 | V | N | 144 | ⋯ | ⋯ | 0.711 | 0.340 | ⋯ | ⋯ | −2.853 | 0.592 | 0.016 | 0.126 | 0.038 | 0.195 |
| 1RXS J150759.8+041511 | X-ray | N | 144 | ⋯ | ⋯ | 0.000 | 0.314 | ⋯ | ⋯ | −1.142 | 0.547 | 0.074 | 0.273 | 0.074 | 0.273 |
| 1RXS J151747.3+652522 | U | N | 200 | ⋯ | ⋯ | 0.222 | 0.132 | ⋯ | ⋯ | −2.569 | 0.205 | 0.006 | 0.076 | 0.007 | 0.086 |
| 1RXS J151747.3+652522 | X-ray | N | 200 | ⋯ | ⋯ | 0.000 | 1.956 | ⋯ | ⋯ | −0.762 | 0.305 | 0.173 | 0.416 | 0.173 | 0.416 |
| 1RXS J153501.1+532042 | B | N | 155 | ⋯ | ⋯ | 1.188 | 0.295 | ⋯ | ⋯ | −4.162 | 0.515 | 0.004 | 0.061 | 0.019 | 0.137 |
| 1RXS J153501.1+532042 | X-ray | N | 155 | ⋯ | ⋯ | 0.275 | 0.370 | ⋯ | ⋯ | −1.479 | 0.645 | 0.083 | 0.289 | 0.124 | 0.353 |
| PG 1553+113 | W1 | Y | 350 | 1.842 | 0.097 | 0.682 | 0.041 | 0.270 | 0.057 | −2.358 | 0.061 | 0.045 | 0.211 | 0.087 | 0.295 |
| PG 1553+113 | X-ray | Y | 350 | 0.888 | 0.196 | 0.835 | 0.105 | 0.349 | 0.027 | −1.449 | 0.066 | 0.314 | 0.561 | 0.479 | 0.692 |
| PKS 2005-489 | V | N | 206 | ⋯ | ⋯ | 0.380 | 0.221 | ⋯ | ⋯ | −3.477 | 0.353 | 0.001 | 0.036 | 0.003 | 0.053 |
| PKS 2005-489 | X-ray | N | 206 | ⋯ | ⋯ | 0.584 | 0.199 | ⋯ | ⋯ | −2.353 | 0.318 | 0.035 | 0.187 | 0.109 | 0.330 |
| PKS 2155-304 | V | Y | 250 | ⋯ | ⋯ | 0.454 | 0.010 | ⋯ | ⋯ | −2.180 | 0.195 | 0.033 | 0.183 | 0.056 | 0.236 |
| PKS 2155-304 | X-ray | Y | 200 | ⋯ | ⋯ | 0.089 | 0.102 | ⋯ | ⋯ | −0.878 | 0.199 | 0.186 | 0.431 | 0.203 | 0.450 |
| 3C 454.3 | W1 | Y | 150 | 0.766 | 0.159 | 0.792 | 0.091 | 0.208 | 0.037 | −1.542 | 0.051 | 0.331 | 0.575 | 0.384 | 0.620 |
| 3C 454.3 | X-ray | Y | 150 | 1.06 | 0.116 | 0.663 | 0.059 | 0.123 | 0.041 | −1.233 | 0.044 | 0.187 | 0.423 | 0.210 | 0.458 |
| H2356-309 | V | N | 300 | ⋯ | ⋯ | 0.133 | 0.162 | ⋯ | ⋯ | −2.505 | 0.334 | 0.005 | 0.070 | 0.005 | 0.077 |
| H2356-309 | X-ray | N | 300 | 1.866 | 0.290 | 0.915 | 0.314 | 0.000 | 0.372 | −2.354 | 0.415 | 0.100 | 0.316 | 0.221 | 0.471 |
| stack | B, V | N | 100 | ⋯ | ⋯ | 0.138 | 0.104 | ⋯ | ⋯ | −2.531 | 0.159 | 0.971 | 0.986 | 0.976 | 0.988 |
| stack | X-ray | N | 200 | ⋯ | ⋯ | 0.240 | 0.100 | ⋯ | ⋯ | −1.331 | 0.154 | 1.049 | 1.024 | 1.049 | 1.024 |

Note. $\Delta t$ notes the sampling interval out to which the curve fit is applied, and the SF and its square root at 30 and 100 days present the source's variability on a timescale of one to three months (a factor to consider when observing variable behavior or mission planning, which takes place on a similar timescale). Fields marked with ⋯ indicate that the curve fit used for that data set is a single fit, rather than a piecewise fit, therefore parameters $k_2$ and xi are not applicable.

**Table 6**
Percentage of X-Ray Observations Exceeding X-Ray Trigger Flux

| Source Name | Filter | $0.9 F_{X,\text{avg}}$ | $1.0 F_{X,\text{avg}}$ | $1.1 F_{X,\text{avg}}$ | $0.9 F_{X,\text{med}}$ | $1.0 F_{X,\text{med}}$ | $1.1 F_{X,\text{med}}$ |
|---|---|---|---|---|---|---|---|
| Mrk 421 | M2 | 49.2 | 42.3 | 36.8 | 55.2 | 49.8 | 43.2 |
| 3C 454.3 | W1 | 46.7 | 39.7 | 35.9 | 56.1 | 49.8 | 42.7 |
| PG 1553+113 | W1 | 51.9 | 45.1 | 41.1 | 54.1 | 50.0 | 44.1 |
| 2MASX J14283260+4240210 | U | 57.3 | 49.7 | 37.4 | 57.3 | 49.7 | 39.2 |
| 3C 273 | V | 61.6 | 42.1 | 27.4 | 67.9 | 47.3 | 27.4 |
| PKS 2155-304 | V | 47.3 | 39.5 | 32.6 | 55.6 | 49.9 | 43.9 |
| Average | | 52.3 | 43.1 | 35.2 | 57.7 | 49.4 | 40.1 |

**Note.** These percentages were calculated without outliers, resulting in median threshold percentage values that are slightly lower than 50%.

1RXS J153501.1+532042.1+532042 (*B*), H2356-309 (*V*), PG 1437+309 (*V*), PKS 2005-489 (*V*), and Ton 116 (*B*).

The sources that are within the stack are fainter than those with more observations, and overall errors (positive and negative) are greater for the stack than for its individual sources. The increased error results from both the effect of propagating errors for more than one source and the fact that the stacked objects had larger than average individual errors in their raw data.

## 4. General Characteristics of Variability

We defined an object to be "variable on short timescales" if the flux changed by least 20% within 50 days during any part of the object's time series. Sources that are categorized as "not variable on short timescales" either exhibit changes in flux that do not exceed 20% on timescales up to 50 days or have too few or widely spread observations to draw conclusions about their short-term behavior. 13/20 of the sources (including the stack of sources) are classified as "variable on short timescales" in the X-ray, and 10/20 in the UV/optical, by this definition.

Most data sets contained observations that are spread out over hundreds, or thousands, of days. However, occurrences of X-ray flux variations that exceed 20% within a period of 50 days were prevalent in most sources that contained observations at such short intervals of time. As each data set contains one measurement of X-ray flux density and one of UV/optical flux for each observation, the arrays containing the respective fluxes are the same length and can be plotted as pairs (illustrated in Figures 1 and 2).

The UV/optical flux variation rarely deviates by more than a factor of five either above or below the calculated average flux. UV/optical flux densities generally remain within 15% of their previous value, although the most drastic short-term variation in UV/optical flux appears in source 3C 454.3, which exhibited a 400% outburst in UV/optical flux within 30 days. Table 3 illustrates the mean, median, and range of flux densities for each data set.

When sources are brighter than average, their X-ray and UV/optical flux are more likely to vary during short periods of time (less than 30 days), as bursts of radiation are short-lived and typically level out within 20–30 days. This phenomenon is more readily seen in a particular group of sources (comprising 2MASX J14283260+4240210, 3C 454.3, Mrk 421, and PG 1553+113) that has many observations over a long period of time and therefore contain a more complete illustration of variable behavior at high brightness.

In all, 13 out of the 20 sources, including the stack, feature flux changes meeting or exceeding 20% within 50 days, and 10 out of the 20 sources exhibit this behavior in the UV/optical. Bursts of high flux are usually short-lived, and the most documented sources appear to be the most variable ones. The X-ray flux varies to a greater extent than the UV/optical flux in nearly all sources.

## 5. Structure Function Analysis

### 5.1. Introduction

When planning an observations, there is a time difference between choosing a target (based on its brightness) and the date the observation is obtained. For a space mission, this time difference can be weeks to months, so it is important to have a quantitative estimate of the typical flux variation as a function of time separation. One approach is the analysis of an X-ray- or UV/optical-triggered response in flux, and here we use a complementary approach, the structure function (SF), which is a numerical method that allows one to determine the variation between two time differences and its behavior as the time differences become longer. The shape of the SF can help determine the proximity of a source's behavior to certain types of noise, such as red noise, which is observed on QSOs on longer timescales, or white ($1/f$, flickering) noise, which is likely related to intrinsic disk or accretion processes (Manners et al. 2002).

### 5.2. Definition of the Structure Function

The structure function provides the mean difference in the flux densities as a function of the separation in the sampling interval (e.g., Simonetti et al. 1985; Hufnagel & Bregman 1992). The measurement is defined as

$$\text{SF} = \frac{1}{N} \Sigma [a(t) - a(t + \Delta t)]^2 \quad (4)$$

where $\Delta t$ represents the sampling interval in which the flux differences are subtracted, $a$ is the flux at a given sampling interval, and $N$ is the number of data points that are within the interval.

We set the standard sampling interval to one day with the time-delay data. Small sampling intervals of 0.1, 0.3, and 0.5 days were added for sources with multiple time-delay values that are below one day. These sources, Mrk 421, 3C 454.3, PG 1553+113, 2MASX J14283260+4240210, 3C 273, and PKS 2155-304, have an overall greater number of observations that are closer together in time. For each $\Delta t$ interval, we averaged the sum of the squares of all flux differences that are within the interval to yield the structure function.

**Table 7**
Notes on Individual Objects: X-Ray Triggering

| Object | $0.9F_{X,avg}$ | $1.0F_{X,avg}$ | $1.1F_{X,avg}$ |
| --- | --- | --- | --- |
| Mrk 421 Figure 2 | Typical 40%–80% advantage relative to the mean | Advantage appears to be a bit larger but remains in the 60%–80% range. The $1.0F_{X,avg}$ trigger produces a fairly consistent 10%–12% advantage from 0 to 100 days over the $0.9F_{X,avg}$ trigger. | On average 70% brighter from 0 to 25 days and 77%–79% brighter from 26 to 85 days compared to untriggered flux. $1.1F_{X,avg}$ triggered flux appears to be 10%–15% brighter than the $1.0F_{X,avg}$ triggered flux from 0 to 100 days. |
| 3C 454.3 Figure 2 | 15%–20% advantage from 0 to 110 days with a fairly consistent 40%–60% increase over the 300 day period | Advantage is similar to $0.9F_{X,avg}$ trigger but is about 5%–8% brighter. | 30%–31% brighter than the untriggered flux. Advantage is consistently around 5%–7% brighter from 0 to 300 days than $1.0F_{X,avg}$. |
| PG 1553+113 Figure 2 | Advantage appears to climb from 50% to 70% out to 100 days, reaching a varying advantage about a mean of 70%–74% beyond 150 days and slowly decreasing out to 300 days. | Increases from 55% to 85% from 0 to 110 days. On average 5%–7% brighter than $0.9F_{X,avg}$ triggered flux from 0 to 100 days before converging around 150 days. | Increasing 57%–90% advantage from 0 to 100 days. Varies around an average 75%–80% advantage from 150 to 240 days before decreasing slightly out to 300 days. |
| 2MASX J14283260+4240210 Figure 2 | Advantages of about 20% from 0 to 45 days (peak around 42% at 10 days), after which the advantage decreases steadily from 46 to 110 days and appears modest beyond 175 days, with large variation due to the sample size. | Consistently around 20% brighter than untriggered flux. Around 5% brighter than the $0.9F_{X,avg}$ trigger from 0 to 110 days. | Produces a similar advantage profile to that of the $0.9F_{X,avg}$ and $1.0F_{X,avg}$ triggers but remains at a 25%–30% increase from 21 to 100 days. There is an advantage of 5%–10% over the flux produced by a $1.0F_{X,avg}$ trigger before converging to $0.9F_{X,avg}$ fluxes around 120 days. |
| 3C 273 Figure 2 | Modest advantages averaging around 7%–10% from 0 to 150 days. | Advantage is around 15% from 0 to 25 days and 5%–8% from 45 to 77 days. Produces an advantage over a $0.9F_{X,avg}$ trigger that is around 5%, varying between 0% and 10%. | Advantages vary between 15% and 45% from 0 to 100 days (on average 30%). Advantages of >40% occur at 55–58, 74, 82–87, 97–100, 135–150, and 225 days, but results at larger $\Delta t$ may be due to the limited number of observations in these bins. |
| PKS 2155-304 Figure 2 | A widely varying but 55%–57% average advantage decreases to 20% from 0 to 300 days. | Produces a 10%–15% advantage over the $0.9F_{X,avg}$ trigger, converges around 130 days. | Similar advantage to the $1.0F_{X,avg}$ trigger. |
| Stack Figure 2 | Advantage is around 20%–28% from 0 to 60 days, gradually decreases with time. | Similar to the $0.9F_{X,avg}$ trigger, but is approximately 5%–10% brighter from 0 to 300 days. | Predicted to result in 35%–40% advantages, but triggering is not as useful from 0 to 300 days, approaching the advantages yielded by the $1.0F_{X,avg}$ trigger instead. |

### 5.3. Structure Function Error Propagation

We utilized the standard deviation in the mean as the error of the structure function for each $\Delta t$. The first steps of the calculation are defined as

$$\Delta a = a(t) - a(t + \Delta t) \quad (5)$$

$$\Delta a_{\mathrm{err}} = \sqrt{a(t)_{\mathrm{err}}^2 + a(t + \Delta t)_{\mathrm{err}}^2} \quad (6)$$

$$\Delta a_{\mathrm{err}}^2 = \Delta a^2 \sqrt{2\left(\frac{\Delta a_{\mathrm{err}}}{\Delta a^2}\right)^2} \quad (7)$$

where $\Delta a$ is the flux difference between two observations that are within the same sampling interval. This process is repeated for each flux difference within one $\Delta t$ and then for each sampling interval.

One takes the standard deviation of the individual errors within one $\Delta t$, where there are $n$ component values for a $\Delta t$. The standard deviation in the mean, or uncertainty for one particular sampling interval, is calculated in the usual way:

$$\sigma_{\mathrm{SF,mean}} = \frac{\sigma_{\Delta a_{\mathrm{err}}^2}}{\sqrt{n-1}}. \quad (8)$$

### 5.4. Curve-fitting the Structure Function

Fitting the structure function is carried out parametrically in the $\log(\mathrm{SF})$–$\log(\Delta t)$ space, assuming a log-normal distribution. Single-segment linear fitting (power law) is appropriate and utilized for sources with fewer than 100 observations, but most objects require a piecewise dual-segment linear fit. The more complicated fit allows one to determine a turnover point ($xi$) in which the SF flattens and exhibits behavior that switches from one noise pattern to another (i.e., red noise to white or "flickering" noise). Two variations of the simple linear equation of the form

$$\log(\mathrm{SF}) = m \log(\Delta t) + b \quad (9)$$

are utilized to fit our data for the dual-segment linear fit. Before the intersection point of the two lines ($xi$), the data are fitted to the first equation:

$$\log(\mathrm{SF}) = k_1 \log(\Delta t) + a_1 \quad (10)$$

where $y$ is the SF at a given $\Delta t$, $k_1$ is the slope of the first line and $a_1$ is the $y$-intercept. The second, flattened line of the piecewise curve fit with slope $k_2$ has a domain that begins at the turnover point ($xi$). The structure function value that corresponds to the $\Delta t$ of the turnover and $xi$ comprise the coordinates of the intersection point $\Lambda$:

$$\Lambda = k_1 \times xi + a_1 \quad (11)$$

$$\log(\mathrm{SF}) = k_2 (\log(\Delta t) - xi) + \Lambda. \quad (12)$$

The four variables $xi$, $k_1$, $k_2$, and $a_1$ are returned from the curve fit function as log values. We set bounds for these parameters to ensure that the values returned are within the sampling interval used to fit the structure function. We limited the value of $xi$ to be between $-1$ and 2.5 (in log-normal space), as the curve fit is applied for $\Delta t$ that are between these orders of magnitude. The quantities $k_1$ and $k_2$ were set to be greater than zero, with $k_2$ limited to be between zero and $k_1$, creating a flatter slope.

In the single-segment linear fits, the only parameters that are determined are $k_1$ and $a_1$, and one equation (below) is utilized to fit the SF data:

$$\log(\mathrm{SF}) = k_1 \Delta t + a_1. \quad (13)$$

The restriction applied is that $k_1$ must exceed zero (an object does not have negative variable behavior). Both fitting routines produce a two-dimensional array whose diagonals provide the variance of the parameter estimates, with the square root of the array diagonals yielding the standard deviation errors of each of the parameters.

### 5.5. Calculation of the Stack Structure Function

The stack comprises 12 objects, all containing 50 or fewer individual observations in the X-ray, $B$, and $V$ wave bands. The structure function for the stack is calculated by assessing the SF for each source within the stack separately before overlaying each scatter plot of the SF in the final figure.

The errors of the stack structure function are calculated by the same method as the other objects. The combined stack is utilized to determine the errors because the SFs of the individual objects do not always contain more than one observation in each $\Delta t$ value with which to calculate $\sigma_{\mathrm{SF,mean}}$.

A linear curve fit is applied to the time-delay and flux arrays of the combined stack to determine the variable behavior of the stack as if it were an individual object. A piecewise curve fit of the stack SF yields no turnover point, so the more complicated fit is not necessary.

### 5.6. Source Selection

The SF and curve fits were calculated for the most populated filter (UV/optical and X-ray) for every source (Figure 4, Table 4). Objects with over 50 individual observations provided the most precise curve fits and the smallest parameter uncertainties. The six sources with the most well-defined SF will be the primary subject of discussion, although the figures for SF calculations on smaller data sets are included. The SF curve fits (Figure 7) of 3C 454.3 (W1 and X-ray), Mrk 421 (M2 and X-ray), PG 1553+113 (W1 and X-ray), 2MASX J14283260+4240210 ($U$ and X-ray), 3C 273 ($V$ and X-ray), and PKS 2155-304 ($V$ and X-ray) include small sampling intervals of 0.1, 0.3, and 0.5 days (or a combination of such intervals).

### 5.7. General Characteristics of Curve Fits

Most of the objects exhibit a piecewise curve fit that increases until it reaches a turnover point, where the slope of the fit decreases. The second slope ($k_2$) remains close to zero, with the average across all SF calculations at 0.18, and it is most often less than half of the first slope ($k_1$). The average $k_1$ value is about 0.61 in the X-ray and 0.57 in the UV/optical with an overall average of 0.59, but this parameter varies between 0.0 and 1.84 across all sources. The curve fits for Mrk 421 (X-ray, Figure 7), 3C 454.3 (W1, Figure 7), PG 1553+113 (X-ray, Figure 7), 2MASX J14283260+4240210 (X-ray, Figure 7), 3C 273 (X-ray, Figure 7), 1RXS J150759.8 +041511 ($V$), S5 0836+71 (X-ray), 1RXS J111706.3 +201410 ($V$ and X-ray), 1ES 1028 (X-ray), H2356-309 (X-ray), PG 1437+389 ($V$ and X-ray), Ton 116 ($B$ and X-ray), and 1RXS J153501.1+532042 ($B$) have $k_1$ values that are between 0.7 and 1.3 (or within 30% of 1.0), which is consistent with red noise, which has a power-law index of 1.0.

**Table 8**
Number of Days in Which the Advantage Produced by an X-Ray Trigger Decreases by a Factor of Two

| Object | $0.9F_{X,avg}$ | $1.0F_{X,avg}$ | $1.1F_{X,avg}$ |
|---|---|---|---|
| Mrk 421 | >500 | >500 | >500 |
| 3C 454.3 | 450 | 400 | 310 |
| PG 1553+113 | 400 | 350 | 325 |
| 2MASX J14283260+4240210 | 90 | 110 | 75 |
| 3C 273 | 50 | 90 | 78 |
| PKS 2155-304 | 120 | 95 | 84 |
| Stack | 75 | 65 | 45 |

**Note.** The advantage for Mrk 421 does not appear to decrease to half of its original value during its total observing timescale.

A turnover point in either the UV/optical or X-ray occurs before 100 days in 10 out of the 12 sources that utilize a piecewise curve fit, and it occurs before 10 days in eight of the 12. Objects PKS 2155-304 (X-ray, Figure 7) and the stack ($B + V$ and X-ray, Figure 7) exhibit white noise, or flickering, where the fit is mostly flat ($k_1 \sim 0$), and do not feature a distinct turnover point. Objects 1RXS J022716.6+020154 (X-ray), 1RXS J150759.8+041511 (X-ray), 1RXS J151747.3+652522 ($U$ and X-ray), and PKS 2005-489 (X-ray) also have power-law indices that are indicative of white noise, but the limited number of observations does not indicate a high degree of precision in their power-law indices.

Across all sources (Table 5), one observes that there are significantly smaller SF values for UV/optical data than for X-ray data. X-ray SF calculations for each sampling interval are typically 2–4 times larger than the UV/optical SF calculations for $\Delta t < 300$ days. Beyond these $\Delta t$, the X-ray SF values become closer to those in the UV/optical bands. A notable exception is object 1RXS J022716.6+020154, whose UV/optical values vary between $0.3\times$ and $4\times$ the X-ray values. Standard deviation error (in the mean) is typically 20%–80% of the average SF for each sampling interval at both the UV/optical and X-ray wavelengths. A complete figure set of the six most documented sources' curve fits (including the stack) is available to view in the online journal (seven images).

To conclude, the six most observed sources and the stack were fitted to piecewise linear or single-segment linear curve fits to determine the variability behavior of the source's flux and the turnover point at which the source's SF flattened and switched between noise patterns. Of the six sources and the stack, four have turnover points less than 10 days, and PKS 2155-304 and the stack appear to "flicker" in the X-ray and UV/optical. Of all objects, Mrk 421 (X-ray, Figure 7), Ton 116 (X-ray), and 1RXS J111706 ($V$) have the closest $k_1$ values to that of red noise (1.0), and the average value for this parameter is 0.61 in the X-ray and 0.57 in the UV/optical across all 19 sources and the stack. The UV/optical SF values are significantly smaller than those of the X-rays until $\Delta t \sim 300$ days for most sources.

### 5.8. Mission Planning Implications

The X-ray fractional flux variation (square root of the SF) can provide insight when mission planning because its value at a specific sampling interval allows one to infer the variation of the source's flux at a given $\Delta t$. These variation predictions will be more accurately obtained with a greater number of observations, though the flux of a source may change drastically after selecting it for observation. The structure function and variation of each source at 30 and 100 days are shown in Table 5, which characterizes the difference between a monitoring observation and a potential mission observation, respectively.

The sources with the greatest SF values and variation at 30 days are 1RXS J022716.6+020154, Mrk 421, 1RXS J122121.7+301041, and 1RXS J150759.8+041511 (all X-ray), with variations that are within 30% of a 1.0 normalized flux unit, indicating that the X-ray flux varies, on average, by this amount at 30 days from the initial observing time. These sources are followed by 3C 454.3 (X-ray and W1), the stack, 1RXS J111706.3+201410, and PG 1553+113 (the last three in the X-ray), with variations of 0.5–0.7 normalized flux units.

To mitigate the presence of potentially low flux measurements, we consider that a strategy using an X-ray or UV/optical trigger (instead of all of the data) to observe a source will more likely yield overall greater luminosities and improve mission planning.

## 6. Advantages to Triggering Observations

### 6.1. UV/Optical and X-Ray Triggering Procedure

The structure function calculations show that, relative to an observation prior to constructing an observing schedule (at time $t_0$), the object can dim enough that the nominal goal for the signal-to-noise ratio is not achieved. This approach does not make use of the object flux at $t_0$ (an untriggered strategy), but we find that one can improve the result by committing to an observation only if the flux at $t_0$ is above some flux level, taken as a fraction of the mean flux (the triggered strategy). This can be a valuable procedure because, in mission planning, one must set up an observing schedule days or weeks in advance, therefore necessitating a reasonably accurate flux forecasting method to determine whether a source is bright enough to observe.

We utilized our minimum-triggering procedure with binned data sets exceeding 50 observations, including the stack that we created. We created threshold fluxes that are specific percentages of the calculated mean. We set these thresholds as $0.9F_{avg}$, $1.0F_{avg}$, and $1.1F_{avg}$ to compare with the entire binned data set to take into account the most probable range of fluxes that would be documented in an observing run.

For each of these limits, we took the average of the X-ray fluxes in each bin that met or exceeded the calculated threshold flux, or averaged all of the fluxes within the bin if there were no fluxes exceeding the trigger. The result is an array of calculated averages that correspond to each bin. Higher and lower threshold percentages were also considered, but increased triggering values greatly diminished the percentage of the data set with which one can utilize triggering.

The penalty for using a triggered response is that the trigger criteria will only be satisfied a fraction of the time. Thus, one would monitor several targets with the goal that at least one will lie above the trigger value to fill out an observing program. A trade-off exists between the trigger value and the fraction of observations exceeding the trigger for the X-ray observations of 3C 454.3, where 50% of the observations lie below $0.88F_{avg}$ and 75% lie below $1.31F_{avg}$ (Figure 5).

Using a trigger as a gateway for observing requires that multiple objects be monitored at the scheduling stage, with the requirement that one object will exceed the trigger a large

Table 9
Notes on Individual Objects: UV–Optical Triggering

| Object | $0.9F_{\rm O,avg}$ | $1.0F_{\rm O,avg}$ | $1.1F_{\rm O,avg}$ |
|---|---|---|---|
| Mrk 421 Figure 3 | Using a UV/optical trigger leads to no advantage. | Similar to the $0.9F_{\rm O,avg}$ trigger, instead remaining entirely at a disadvantage from 0 to 100 days. | 0.5% disadvantage at 5 days increases to 22% at 50 days and continues to increase past 100 days. Similar to the $0.9F_{\rm O,avg}$ and $1.0F_{\rm O,avg}$ triggers, though it appears more "flattened." |
| 3C 454.3 Figure 3 | 20% advantage from 0 to 80 days increases from 80 to 180 days to around a 70% advantage but deceases to the mean around 300 days. | 25% advantage from 0 to 70 days. Consistently 7%–9% brighter than the $0.9F_{\rm O,avg}$ trigger from 0 to 100 days but begins to overlap and decrease beyond 125 days. | On average 34%–37% advantage from 0 to 70 days. Consistently 2%–5% brighter than the $1.0F_{\rm O,avg}$ trigger from 0 to 100 days, afterwards becoming less stable and exhibiting large decreases in triggered flux. |
| PG 1553+113 Figure 3 | Around 20% brighter from 0 to 160 days, afterwards increasing to advantages that are around 25%–27%. Figure 3 illustrates a that this advantage goes away by 100 days. | Increases from 40% to 60% from 0 to 100 days and decreases to remain around 35%–41% brighter out to 300 days. 10%–23% advantage over the $0.9F_{\rm O,avg}$ trigger is apparent from 0 to 80 days before decreasing beyond 175 days. | Advantage increases from 60% at 27 days to those exceeding 80% beyond 66 days (extreme at about 180 days). On average 15% brighter than the $1.0F_{\rm O,avg}$ trigger from 0 to 300 days. |
| 2MASX J14283260+4240210 Figure 3 | Average 7%–8% advantage increases up to 12% a few times after 70 days. | Advantage stays around 15% from 17 to 100 days, varying regularly between 10% and 25% advantage every five days. Produces a varying 3%–15% advantage over the $0.9F_{\rm O,avg}$ trigger from 5 to 85 days before converging around 130 days. Flux is difficult to predict beyond 175 days and reaches advantages alternating between 0% and 55%. | Advantage varies between 10% and 35% from 17 to 100 days but tends to center around 20%. Illustrates large advantages at 92 and 101 days but quickly decreases and converges to the other triggers. |
| 3C 273 Figure 3 | No advantage from 0 to 300 days. | Yields a 5%–10% advantage from 0 to 30 days and 2%–5% from 46 to 82 days. Produces both advantages and disadvantages exceeding 20% beyond 150 days. | Advantages vary between a 20% decrease (146 days) and 25%–30% increase (82 and 130 days), but there does not appear to be a credible advantage at large $\Delta t$. |
| PKS 2155-304 Figure 3 | Advantages vary between −30% and 80% (2 days) but remain around 30% from 20 to 130 days, decreasing to the mean by 300 days. | Advantage varies between −30% and 160% and varies between a 20% and 80% increase from 60 to 160 days. About the same as the $0.9F_{\rm O,avg}$ trigger beyond 160 days, decreasing to the mean. | Similar to the $1.0F_{\rm O,avg}$ trigger with a more pronounced decrease in resulting flux. |
| Stack Figure 3 | Modest brightness advantage averages 5% from 0 to 300 days, with occasional increases between 20 and 55 days up to approximately 11%. | Produces fluxes in the same time range that are 10%–30% brighter than the untriggered flux. Creates disadvantages up to 30% beyond 70 days. | Does not produce brighter fluxes for a majority of the 300 days timescale, varying widely between −60% and 60% advantages. |

**Note.** Flux brightness advantages produced by each UV/optical trigger in comparison to untriggered X-ray flux (averaged into four-day bins). Summarizes the results in Figure 3.

fraction of the time. As an example, we consider a total sample of 20 targets distributed around the sky, with five that are at optimum sky position for the telescope and are monitored at the moment of making a scheduling decision. We can apply binomial probabilities and calculate that for a trigger excluding 50% of the fluxes (five trials), one or more sources will exceed the trigger 98% of the time. For a stricter trigger that requires the flux to be in the upper 25% of the fluxes, this rate of success drops to 76%, This indicates a scheduling failure about one-quarter of the time, which is probably an unacceptable situation. For this reason, we only consider triggers where the rate of success lies in the 90%–98% range, and this roughly corresponds to flux trigger values of $0.9F_{\mathrm{X,avg}}$–$1.1F_{\mathrm{X,avg}}$, although we examine this below more carefully because the variability behavior is different among targets.

We will compare the resulting, averaged fluxes that are produced by X-ray triggers and those that are produced by UV/optical triggers to determine which method is ideal for triggering an X-ray observation. While it is probable that a brighter X-ray triggering threshold will yield overall brighter X-ray flux averages (increasing the threshold will raise the average, and lower fluxes within the bin will be eliminated), the advantage of utilizing a UV/optical trigger is not as guaranteed.

We rebinned the data into larger intervals because it is often the case that there are few flux differences in one-day intervals. We set a bin size of four days in data sets larger than 20 observations from this time-offset array. This span of time allowed us to adequately simplify our data for threshold. This binning is applied to the time-delay triggering analysis, where a resolution of less than one week is adequate. By binning, there are enough measurements in a bin to apply Gaussian statistics, rather than Poisson statistics, which are optimally used in discrete cases and are more difficult to analyze for this purpose.

*6.1.1. Error Propagation of Triggering Procedure*

We utilize the first and third quartiles to examine the spread of the data in each bin. The asymmetric vertical error bars comprise two different values. The section of the error bar that is above the binned and averaged flux, $\Delta \bar{F}_{\mathrm{bin,upper}}$ is defined as

$$\Delta \bar{F}_{\mathrm{bin,upper}} = Q_3 - \bar{F}_{\mathrm{bin}} \qquad (14)$$

where $Q_3$ is the third quartile of the flux measurements in the bin. The lower section of the error bar that is below the binned and averaged flux, $\Delta \bar{F}_{\mathrm{bin,lower}}$, is defined as

$$\Delta \bar{F}_{\mathrm{bin,lower}} = \bar{F}_{\mathrm{bin}} - Q_1 \qquad (15)$$

where $Q_1$ is the first quartile of the flux measurements in the bin. This method is utilized in our investigation to measure the spread of fluxes in each bin instead of the standard deviation in the mean because the former procedure shows us the different flux ranges above and below the calculated average and does not assume that the underlying distribution of flux measurements within a bin is Gaussian.

*6.2. Results of X-Ray Triggering*

We present results with triggers that are $0.9F_{\mathrm{X,avg}}$, $1.0F_{\mathrm{X,avg}}$, and $1.1F_{\mathrm{X,avg}}$ in each data set for X-ray and UV/optical triggering. Section 6.4 examines and compares such trends resulting from each of the X-ray and UV/optical triggering techniques, followed by an analysis of the results as factors for selecting observing durations.

The binned and averaged flux of each source remains close to the mean calculated flux with no trigger, rarely deviating more than 20% above or below this calculated mean. Untriggered fluxes tend to vary more from their previous value with increasing $\Delta t$ until a point is reached (usually occurring before 50 days) at which the flux variation at a given time remains the same or becomes difficult to predict (flickers). This behavior occurs in nearly all objects exceeding 50 individual observations, which is illustrated in our structure function analysis.

The advantages of X-ray triggering vary widely by source. As a whole, the $1.1F_{\mathrm{X,avg}}$ trigger yields the greatest increase in the resulting X-ray flux (compared to no trigger). Selecting increasingly higher trigger fluxes from $0.9F_{\mathrm{X,avg}}$ to $1.1F_{\mathrm{X,avg}}$ produces consistently higher resultant fluxes when observing an object while it is bright. However, an object exceeds a trigger for only a fraction of the time, so there are fewer trigger opportunities for another higher flux. A triggering value is in the range $0.9F_{\mathrm{X,avg}}$–$1.1F_{\mathrm{X,avg}}$, because it will lead to successful triggers around 35%–52% of the time while avoiding poor observational outcomes in the absence of triggering. Table 6 illustrates the percentage of observations exceeding these values (and their median counterparts) for the six objects with the best data (Figure 5).

The advantage to raising a trigger's percentage of the average flux is more evident in well-documented sources (over 100 data points). 3C 454.3 (Figure 2) exhibits similar X-ray triggering behavior to Mrk 421 (Figure 2) and PG 1553+113 (Figure 2). Across all filters, setting a trigger that is $0.9F_{\mathrm{X,avg}}$ yields a brightness advantage that is fairly consistent from 0 to 30 days when compared to fluxes produced by no trigger, though the exact percentages vary among the three sources. This advantage continues up to 100 days.

Sources 2MASX J14283260+4240210 (Figure 2) and PKS 2155-304 (Figure 2) each exhibit similar brightness advantages with respect to each trigger, though their individual percentage advantages over the untriggered data are different. In the former object, raising the X-ray observing trigger from $0.9F_{\mathrm{X,avg}}$–$1.0F_{\mathrm{X,avg}}$ and $1.0F_{\mathrm{X,avg}}$–$1.1F_{\mathrm{X,avg}}$ both produce an average 5% increase in resulting flux, with the $1.0F_{\mathrm{X,avg}}$–$1.1F_{\mathrm{X,avgg}}$ trigger increase generating a slightly higher (7%–9%) brightness advantage from 0 to 24 days. PKS 2155-304 is a highly variable source that appears to flicker with time, but each trigger increase typically yields a varying 0%–10% flux advantage.

Table 7 describes the flux brightness advantages produced by each X-ray trigger in comparison to untriggered X-ray flux (averaged into four-day bins) and summarizes the results in Figure 6. The complete figure set (seven images) can be viewed in the online journal.

Overall, increasing the trigger percentage of the average X-ray flux produces advantages that are emphasized in well-documented sources and less evident in those that flicker or have fewer observations. We set triggering values of $0.9F_{\mathrm{X,avg}}$, $1.0F_{\mathrm{X,avg}}$, and $1.1F_{\mathrm{X,avg}}$ to produce higher resulting fluxes without sacrificing too many observing opportunities; these threshold values produce successful triggers that occur 35%–52% of the time. The highest X-ray flux advantages appear in Mrk 421 (Figure 4), 3C 454.3, PG 1553+113, and PKS 2155-304 (Figure 6), though the latter features irregular flux behavior

Table 10
Number of Days in Which the Advantage Produced by a UV/optical Trigger Decreases by a Factor of Two

| Object | $0.9F_{O,avg}$ | $1.0F_{O,avg}$ | $1.1F_{O,avg}$ |
|---|---|---|---|
| Mrk 421 | ... | ... | ... |
| 3C 454.3 | 260 | 240 | 145 |
| PG 1553+113 | 135 | 115 | 100 |
| 2MASX J14283260+4240210 | 45 | 61 | 61 |
| 3C 273 | 14 | 21 | ... |
| PKS 2155-304 | 55 | 41 | 41 |
| Stack | ... | ... | ... |

**Note.** One can use a UV/optical trigger for about half of the sources, but it is better to use an X-ray trigger in nearly all cases.

from 0 to 300 days. Mrk 421 appears to "sustain" the advantage produced by $0.9F_{X,avg}$, $1.0F_{X,avg}$, and $1.1F_{X,avg}$ triggers for the longest period of time, followed by other bright, variable sources such as 3C 454.3 and PG 1553+113 (Table 8).

### 6.3. Results of UV/Optical Triggering

A brighter UV/optical trigger does not yield as great an advantage between thresholds as X-ray-triggered results, although it appears as if a brighter UV/optical trigger yields overall higher flux averages than a less bright UV/optical trigger (Tables 9 and 10). Mirroring the X-ray threshold triggering procedure, we also assume that the observing period will occur within 60 days of the last known monitoring observation, but larger $\Delta t$ of up to 100 days are also illustrated.

These advantages vary extensively by source. In general, a $1.0F_{O,avg}$ trigger yields X-ray fluxes that are typically 5%–10% greater than those resulting from a $0.9F_{O,avg}$ trigger, though some data sets have too few data points or possess flux advantages that are not discernible from the results of the previous trigger. The largest increases (76% and 81%) are observed in the $V$ band of source 1RXS J122121.7+301041 at roughly 50 and 100 days, respectively. Increasing the trigger to $1.1F_{O,avg}$ outputs binned averages that are modestly brighter than those triggered by $1.0F_{O,avg}$ and to roughly the same magnitude as the previous trigger increase, if not slightly lower at 3%–10% from 0 to 100 days.

Unlike the X-ray-triggered results, some increases in UV/optical triggering flux yield lower flux measurements, which indicates that the UV/optical and X-ray fluxes do not always positively correlate. Mrk 421, 3C 454.3, PG 1553+113, 2MASX J14283260+4240210, 3C 273, PKS 2155-304, and the stack feature at least one decrease in resulting flux that exceeds 20% any time from 0 to 100 days for either triggering increase. Though increasing the UV/optical observing threshold value generally yields a higher X-ray flux with each increase, it comes at the cost of fewer data points. Returning to the six most well-documented sources and the stack, Figure 3 illustrates the average X-ray flux resulting from each UV/optical trigger and the advantages of setting this trigger at a higher percentage of the average UV/optical flux. The left panels illustrate the binned and averaged flux for the first 100–500 $\Delta t$. The green vertical lines represent the error of each bin for the $1.0F_{O,avg}$ triggered data. The right panels compare the binned and averaged fluxes produced by each triggering value and the untriggered flux. The advantage is calculated by dividing the flux yielded by the trigger by the untriggered flux value for the same $\Delta t$. The complete figure set (seven images) is available in the online journal.

To conclude, we find that a brighter UV/optical trigger produces overall higher X-ray flux averages across the most documented objects. However, this is not the case in all objects, and some, including Mrk 421 and 3C 454.3, sometimes display an anticorrelations between their X-ray and UV/optical fluxes. Also, triggers set by UV/optical data are less successful than those set by using X-ray monitoring observations Much like the results of X-ray triggering, the UV/optical-triggered findings indicate that the advantages of raising the trigger are not as evident for objects that flicker or contain fewer observations, especially as increasing the trigger threshold comes at the cost of fewer data points.

### 6.4. Comparing Trigger Wavelengths in Terms of Flux Forecasting

We observed trends in the data sets that confirmed an overall advantage to X-ray triggering over UV/optical triggering upon the creation of our bins and application of the threshold method. Sources that exhibit a UV/optical triggering advantage over an X-ray trigger are defined as yielding a higher average flux when this UV/optical trigger was used, in comparison to an X-ray trigger of the same percentage of the average X-ray flux.

We found that out of the 51 data sets, five illustrated a higher optically triggered flux than X-ray-triggered averages (for over half of the time period of 0–100 days), seven exhibited a temporary optically triggered advantage, two had too few observations to apply binning and triggering techniques, and the rest (37 data sets) did not yield a higher optically triggered flux for each 10% increase in the triggering percentage (Table 11).

3C 454.3 ($B$, M2, $U$, W1, W2) is the object that possesses this advantage more than 50% of the time from 0 to 100 days, indicating that there is a good chance that the X-ray flux will remain above average for at least 20 days using UV/optical triggers. It is worth noting that this source is particularly bright in X-rays and has been well documented for over 10 yr, with data sets ranging from 294 to 356 individual observations.

Objects with more observations are more likely to have a correlation between X-ray and UV/optical fluxes, positive or negative. Our data suggest that accurate measurements are more likely to lead to strong correlations; large uncertainties in the observations of smaller data sets yielded correlations that were either poor or indeterminate. Although an increase in a UV/optical observing threshold is more likely to be paired with an increase in overall X-ray flux, the result is less bright than an X-ray-triggered flux in nearly all cases.

Some objects do not appear to fit into either group and exhibit optically triggered advantages (for $1.0F_{O,avg}$ or $1.1F_{O,avg}$ triggers) sporadically or for short periods of time for under half of the timescale. These sources are 1RXS J122121.7+301041 (W1), 2MASX J14283260+4240210 ($B$, $U$, W1, W2), 3C 273 (M2, $U$), and PKS 2155-304 ($V$). In these instances, the optically triggered flux tends to yield an inconsistent, modest advantage (on average 5%–10%) and features one or two large advantages of up to 25%, usually with the $1.1F_{O,avg}$ UV/optical trigger between 40 and 100 days.

### 7. Discussion and Conclusions

We obtained observations from 19 individual objects and up to six filters/wave bands over the period from 2005 March to 2020 September. We review the variability results, and when

**Table 11**
Ratio Comparison of X-Ray Flux Brightness Produced by X-Ray and UV/optical Triggers

| Object | $0.9F_{\text{avg}}$ | $1.0F_{\text{avg}}$ | $1.1F_{\text{avg}}$ | Conclusion |
|---|---|---|---|---|
| Mrk 421 Figure 8 | For a trigger of $0.9F_{\text{O,avg}}$, we observe an average 38%–40% disadvantage compared to $0.9F_{\text{X,avg}}$ from 0 to 300 days, decreasing from a disadvantage of 30% at 2 days to one that is 50% at 185 days. | The $1.0F_{\text{O,avg}}$ trigger yields X-ray fluxes around a 40%–60% disadvantage to those produced by $1.0F_{\text{X,avg}}$ from 0 to 100 days, but this improves to a 35% advantage from 170 to 210 days before continuing to decrease once more to 300 days. | The $1.1F_{\text{O,avg}}$ trigger produces a similar disadvantage to the $1.0F_{\text{O,avg}}$ UV/optical trigger, but it reaches a maximum disadvantage of around 80% around 175 days before increasing up to a 40% disadvantage from 210 to 230 days. | No matter the triggering value, it is not ideal to utilize UV/optical triggering for this source within the first 100 days of observation, as the X-ray flux produced by the UV/optical triggers is less bright than those produced by X-ray triggers. |
| 3C 454.3 Figure 8 | The $0.9F_{\text{O,avg}}$ trigger yields similar, if not higher, binned and averaged flux to when the trigger is in UV/optical wavelengths, between a 7% disadvantage and a 6% advantage. All three triggers feature disadvantages (maxima around 40%–60%) beyond 120 days. | The $1.0F_{\text{O,avg}}$ trigger yields advantages over $1.0F_{\text{X,avg}}$ that stay around 5% but decrease at 30 and 60 days to more modest advantages of 0%–5%. The greatest UV/optical advantage for this trigger occurs from 21 to 25 days at 10%. Advantage decreases starting at 120 days to a disadvantage around 45% at 160 days. | The $1.1F_{\text{O,avg}}$ trigger yields advantages over $1.1F_{\text{X,avg}}$ that are on average 5% from 34 to 50 days, but there are decreases at 30 and 60 days to a small UV/optical disadvantage of 2%. This trigger is not advantageous nor stable beyond its steep decrease from 120 to 160 days. | If observing 3C 454.3 in the W1 band within 100 days, a $1.0F_{\text{X,avg}}$ (or $1.1F_{\text{X,avg}}$) UV/optical trigger is ideal when the object is bright, as the X-ray fluxes produced by this trigger are most consistently higher than those that are X-ray-triggered. Triggering with UV/optical wavelengths is not recommended beyond 100 days. |
| PG 1553+113 Figure 8 | For a trigger of $0.9F_{\text{O,avg}}$, the flux is about 25% less bright than those produced by an X-ray trigger from 0 to 300 days, decreasing slightly to −35% from 140 to 170 days. | The $1.0F_{\text{O,avg}}$ trigger produces X-ray fluxes that are 10%–17% less bright than the X-ray-triggered fluxes from 0 to 300 days but decreases to −30% from 140 to 170 days. | The $1.1F_{\text{O,avg}}$ trigger yields up to a 5% advantage at 2 and 56–65 days. On average, the $1.1F_{\text{O,avg}}$ trigger is 12% less bright than the $1.1F_{\text{X,avg}}$ trigger from 0 to 100 days. This trigger's fluxes decrease to a disadvantage around 30% at 160 days and hover around −15% out to 300 days. | It is not optimal to use a UV/optical trigger when triggering observations for this object in the W1 band for the first 300 days of observation. However, an increase in triggering flux appears to improve the brightness yielded by a UV/optical trigger in comparison to its X-ray counterpart. |
| 2MASX J14283260+4240210 Figure 8 | Resulting fluxes from $0.9F_{\text{O,avg}}$ are on average 5%–15% lower than those produced by the $0.9F_{\text{X,avg}}$ trigger from 15 to 70 days. All of the UV/optical triggering advantages and disadvantages hover around the same values out to 300 days but increase in variation. | The $1.0F_{\text{O,avg}}$ trigger exhibits similar behavior to the $0.9F_{\text{O,avg}}$ trigger but remains between a 10% disadvantage and 5% advantage from 20 to 200 days. | $1.1F_{\text{O,avg}}$ optically triggered flux varies similarly to the $1.0F_{\text{O,avg}}$ trigger but reaches disadvantages in comparison to $1.1F_{\text{X,avg}}$ from 70 to 220 days that are up to 35%. The largest advantage this UV/optical trigger possesses is a 7% advantage at 74 days. | It is not ideal to use UV/optical triggering for this source within 300 days, as the advantages over the X-ray trigger present are inconsistent and modest (under 10%). |
| 3C 273 Figure 8 | We observe a 4%–7% disadvantage for $0.9F_{\text{O,avg}}$ to X-ray-triggered fluxes from 0 to 100 days, decreasing to greater disadvantages exceeding 20% beyond 140 days. | The $1.0F_{\text{O,avg}}$ trigger, similar to the $0.9F_{\text{O,avg}}$ trigger, remains at a 5% brightness disadvantage that increases to disadvantages up to 45% from 135 to 200 days. | The $1.1F_{\text{O,avg}}$ trigger has a disadvantage that varies the most among the three UV/optical triggers. It varies between a disadvantage of 35% and advantage of 1% (at 45 days), resulting in an overall disadvantage of 13% from 0 to 300 days. | It is not ideal to utilize UV/optical triggering for this source in the V band, but the fluxes yielded by the $0.9F_{\text{X,avg}}$ and $1.0F_{\text{X,avg}}$ UV/optical triggers from 0 to 80 days are not at a very high disadvantage and remain fairly consistent. |

**Table 11**
(Continued)

| Object | $0.9F_{avg}$ | $1.0F_{avg}$ | $1.1F_{avg}$ | Conclusion |
| --- | --- | --- | --- | --- |
| PKS 2155-304 Figure 8 | We observe resulting fluxes from the $0.9F_{O,avg}$ trigger that are on average 13%–15% less bright than those produced by an X-ray trigger from 0 to 300 days. | The $1.0F_{O,avg}$ trigger varies widely between 20% advantages and 60% disadvantages compared to the $1.0F_{X,avg}$ trigger from 0 to 300 days. The average disadvantage posed by this trigger is around 15%–19%. | The $1.1F_{O,avg}$ trigger produces X-ray fluxes that are more variable than those triggered by X-rays. | From 0 to 100 days, the optically triggered flux is at an average disadvantage of 15%. As this source varies greatly on short timescales and exhibits flickering behavior that is difficult to predict, UV/optical triggering is not recommended from 0 to 300 days after the object's initial observation. |
| Stack Figure 8 | $0.9F_{O,avg}$ produces X-ray fluxes that are on average 19%–20% less bright than those produced by a $0.9F_{X,avg}$ trigger. This disadvantage appears to decrease from a disadvantage of around 20% to a more modest value of approximately 9%–10% from 0 to 100 days, remaining close to this value until 300 days and decreasing to a disadvantage around 30% from 150 to 180 days. | In comparison to the $1.0F_{X,avg}$, the $1.0F_{O,avg}$ trigger appears to follow the same trend until 70 days, where its disadvantage increases to beyond 40% and remains around 30% out to 300 days. | It is difficult to statistically compare the results of the $1.1F_{O,avg}$ and $1.1F_{O,avg}$ triggers, as very few flux values exceeded both thresholds for the stack, but comparing the fluxes produced by the two triggers reveals a UV/optical triggering disadvantage that exceeds 50% occasionally from 0 to 30 and 140–300 days. | UV/optical triggering is not ideal for the "object" that the stack represents, though a less bright UV/optical triggering value yields brightness disadvantages that are more moderate and consistent. |

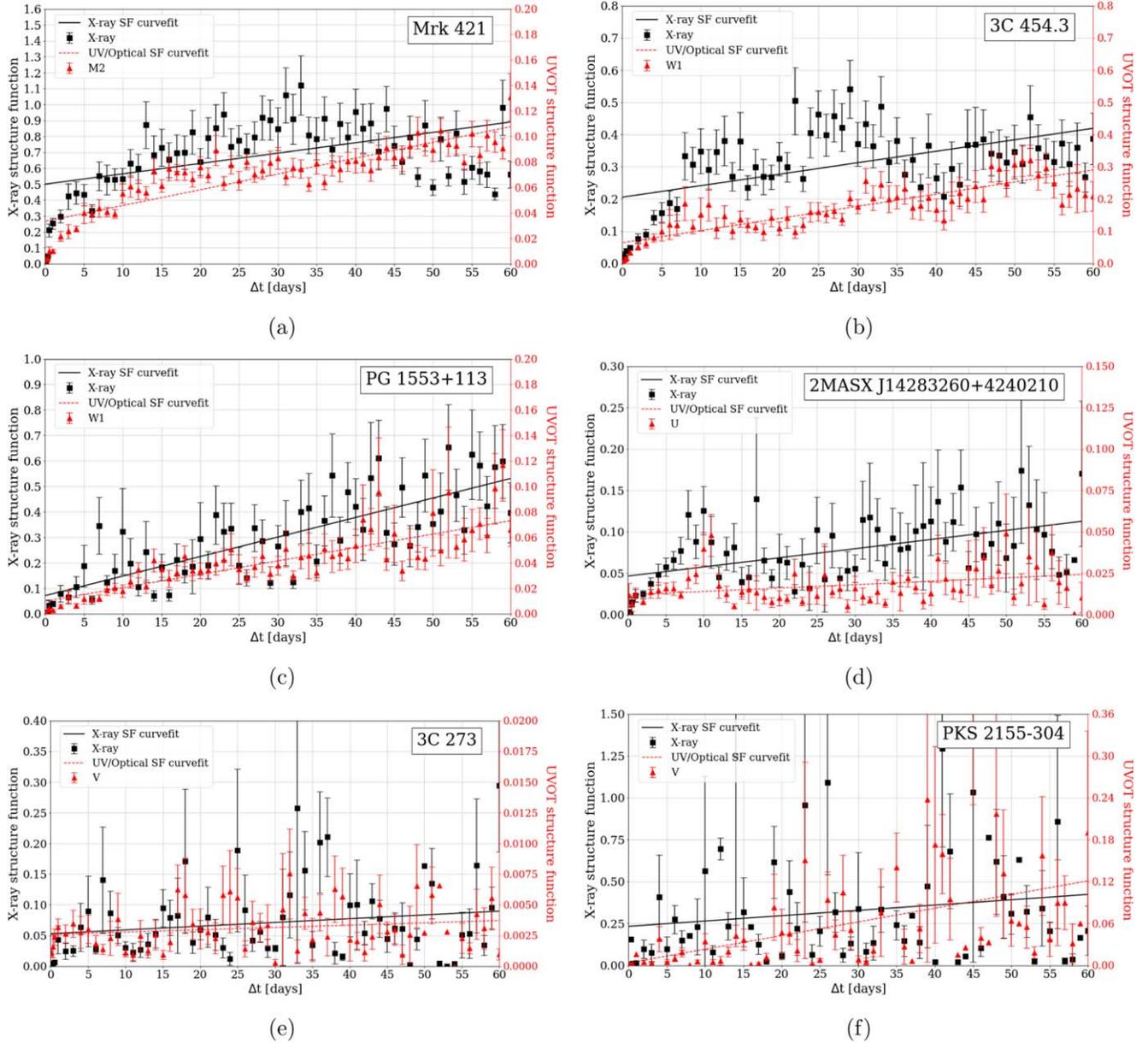

**Figure 3.** Comparison of UV/optical and X-ray structure functions with error bars for the first 60 $\Delta t$ for the six most documented sources This comparison is not illustrated in log–log space, which clarifies the illustration of the first 60 individual SF values and their error bars. The error bars are produced by the first method of uncertainty calculation (standard deviation of points within one $\Delta t$ divided by the square of the number of points minus one). The UV/optical SF of 3C 454.3 is a considerable fraction (40%–50% or exceeding 110% at times) of the X-ray SF in comparison to the other well-observed sources whose UV/optical SF values are typically 10%–30% of the X-ray values.

considering the total range of sampling intervals, we find that nearly all sources exhibit a smaller range of flux variation at UV/optical than at X-ray wavelengths. The two exceptions are 3C 273 (W2 band versus the X-ray band) and 3C 454.3 (U band versus the X-ray band). Our variability results for PKS 2005-489, Mrk 421, and 3C 454.3 agree with those of the recent literature that have utilized the same data from Swift (Raiteri et al. 2011; Aleksić et al. 2015; Kapanadze 2021). The overall anticorrelation between the X-ray and UV/optical fluxes, and power-law spectral densities present in the nonflaring states of Mrk 421 are especially evident, along with the prominent X-ray variability (which follows the UV/optical) of 3C 454.3.

The average range in normalized flux units across all 19 sources is 1.94 for the X-ray band and 0.99 for the UV/optical band, with the best observed objects illustrating the largest ranges in both fluxes (without outliers). The largest ranges are 4.24 normalized flux units in the UV/optical (3C 454.3, $U$ band) and 5.86 in the X-ray band (Mrk 421, M2 filter), while the smallest ranges in the UV/optical and X-ray fluxes are 0.14 normalized flux units (3C 273, $B$ band) and 0.76 (1RXS J153501.1+532042, $B$ filter), respectively.

The rank for the target selection was the average $F_x$ times redshift, where $F_x$ is in the observer's frame (0.5–2 keV) rather than the rest frame. For similar values of the merit parameter, $F_z$, we expect that the more distant objects will be more luminous in the observer's frame.

The flux is proportional to the luminosity divided by the square of the luminosity distance ($d_L$), which equals the product of $(1 + z)$ and the comoving distance. Then, $d_L$ increases somewhat more rapidly than $z$ over our redshift range

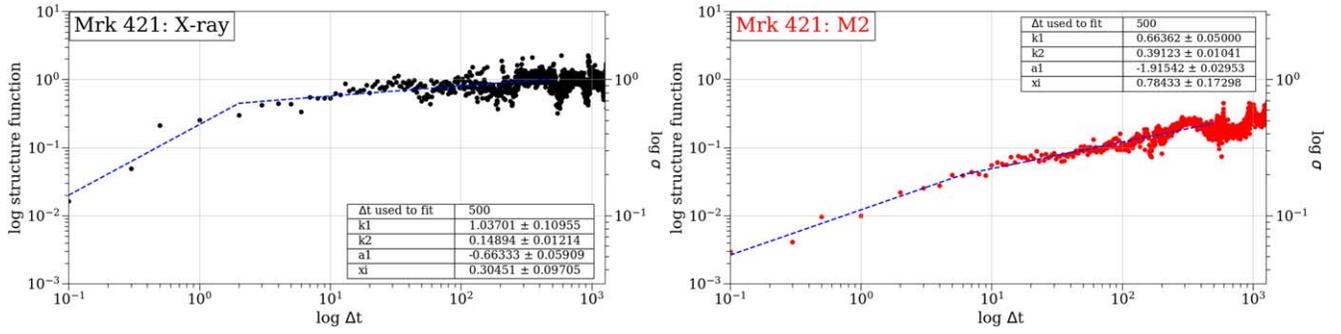

**Figure 4.** Mrk 421 in X-rays (left) and M2 band (right). Both SF analyses utilize small sampling intervals of 0.1, 0.3, and 0.5 days. The X-ray SF features an *xi* of $2.01 \pm 1.25$ days. The $k_1$ power-law index of the X-ray curve fit is the closest of all sources to that of pure red noise preceding the turnover point. The UV/optical SF *xi* of $6.09 \pm 1.49$ days is supported by the small $\Delta t$ SF values. Both the UV/optical and X-ray SF values follow their respective $k_2$ slopes and do not differ from the line by more than one order of magnitude for $\Delta t$ exceeding 500 days. The complete figure set (7 images) is available in the online journal.

(The complete figure set (7 images) is available.)

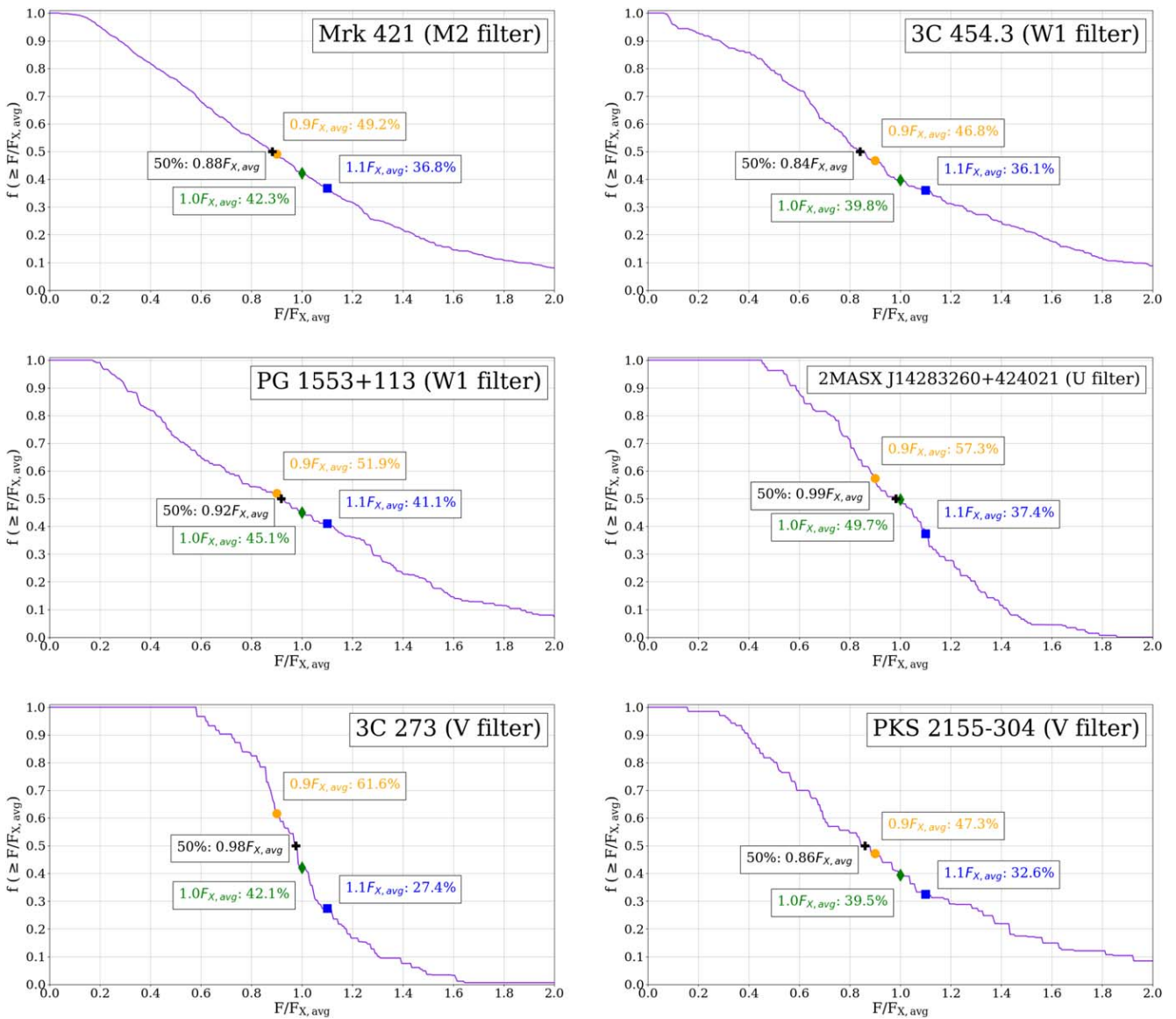

**Figure 5.** Fraction of X-ray observations meeting or exceeding a normalized flux, $F/F_{\rm X,avg}$, for the six objects with the best data. The percentages of an object's X-ray observations that exceed a triggering flux of $0.9F_{\rm X,avg}$, $1.0F_{\rm X,avg}$, or $1.1F_{\rm X,avg}$ are with their respective points. These triggering values allow one to produce higher resulting fluxes without sacrificing too many observing opportunities. Sources 2MASX J14283260+4240210, and 3C 273 have the steepest slopes because they vary the least, and there is a dramatic decrease in the percentage of observations above the triggering flux following a trigger of $1.1F_{\rm X,avg}$.

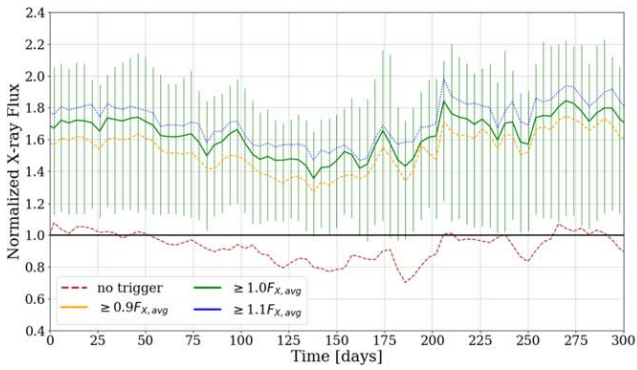 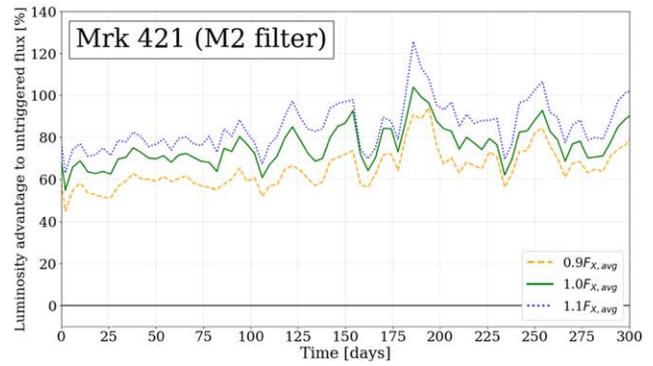

**Figure 6.** Mrk 421 (M2 filter). Quartile errors corresponding to the $1.0F_{X,avg}$ flux are illustrated in green. The flux produced by each trigger appears to remain around $1.4F_{X,avg}$–$1.8F_{X,avg}$ and to decrease with the untriggered flux from 0 to 100 days, later increasing from 100 to 300 days to approximately $1.7F_{X,avg}$ (left). There are decreasing advantages with each increase in trigger flux, as there are fewer fluxes that meet the higher X-ray flux triggers (right). This produces decreasing returns with each trigger increase. Mrk 421 produces some of the highest $0.9F_{X,avg}$ average flux advantages, along with PKS 2155-304 (Figure 2, which has flux advantages exceeding 90%), PG 1553+113 (Figure 2, exceeding 50%), and 1RXS J122121.7+301041 (exceeding 60%). The complete figure set (7 images) is available in the online journal.

(The complete figure set (7 images) is available.)

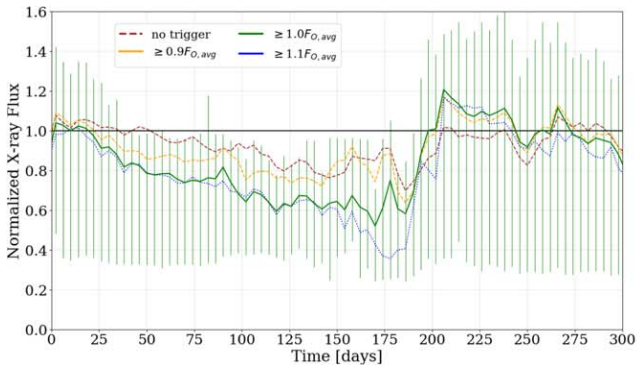 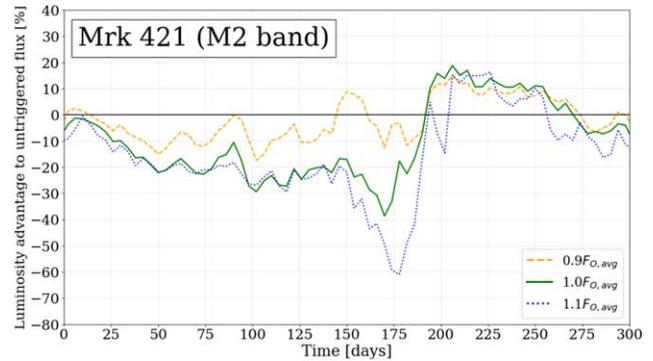

**Figure 7.** Mrk 421 (M2 band). Quartile errors corresponding to the $1.0F_{O,avg}$ flux are illustrated in green. Mrk 421 is unusual in that UV/optical triggering yields almost exclusively lower X-ray fluxes than utilizing no trigger from 0 to 150 days. As the flux of the UV/optical trigger increases, the resulting X-ray flux decreases, indicating a possible negative correlation between the fluxes of the two wavelengths until about 180 days. Though the untriggered flux decreases from 0 to 150 days (left), the optically triggered flux decreases with each increase in triggering flux from $0.9F_{O,avg}$ to $1.1F_{O,avg}$ (right). However, the optically triggered flux experiences a large increase from 180 to 200 days, while the untriggered flux becomes brighter to a smaller extent. The complete figure set (7 images) is available in the online journal.

(The complete figure set (7 images) is available.)

of interest. Consequently, for a flux-limited sample, one might expect that $L$ rises faster than $z$.

The effect is evident in our 19 target systems, where there appears to be a fairly strong correlation between increasing distance and mean luminosity (Figure 9). This correlation appears tighter in the X-ray band, which is expected since there is dispersion in the UV/optical–X-ray flux relationship, and there is greater dispersion in the UV/optical flux (Figure 7).

The distribution of points in Figure 7 does not change when considering BL Lac objects and flat-spectrum radio sources separately, though 3C 454.3 and 3C 273 are relatively bright compared to the other objects. The complete figure set (seven images) is available in the online journal. One might expect that the time variation is slower due to a $(1+z)^{-1}$ cosmological time dilation term and because more luminous objects may be larger. We investigate this hypothesis in both the rest and observer's frames with the 19 objects and correct the values of the structure function, and in turn the variability, in terms of the object's redshift by multiplying every $\Delta t$ value by $1+z$ for the observer's frame. After correcting for the $1+z$ term we find no evidence for more distant objects having slower variations (Figure 7).

There is little evidence for a trend of variability timescale with $z$, but the most variable X-ray object is the closest. We also investigated the possibility of the dependence of variability on luminosity in the observer's frame (Figure 7). We infer that more luminous objects are more variable on shorter timescales (under 20–30 days) than less luminous objects.

The information illustrated in Figure 7 suggests that short-term variability appears to be independent (or positively correlated in the UV/optical band) of average luminosity when investigating flux measurements within their respective wavelengths. This is in conflict with the expectation where the more luminous objects are larger and have slower and smaller variations. We speculate that the result of Figure 7 is due to the high-luminosity sources having a higher Doppler boosting factor due to our closer alignment to the line of sight of the jet. There is suggestive support for this idea in that some of our most luminous sources, such as S5 0836+710, have a higher than average apparent $v/c$ ($\approx 20$) in the very long baseline

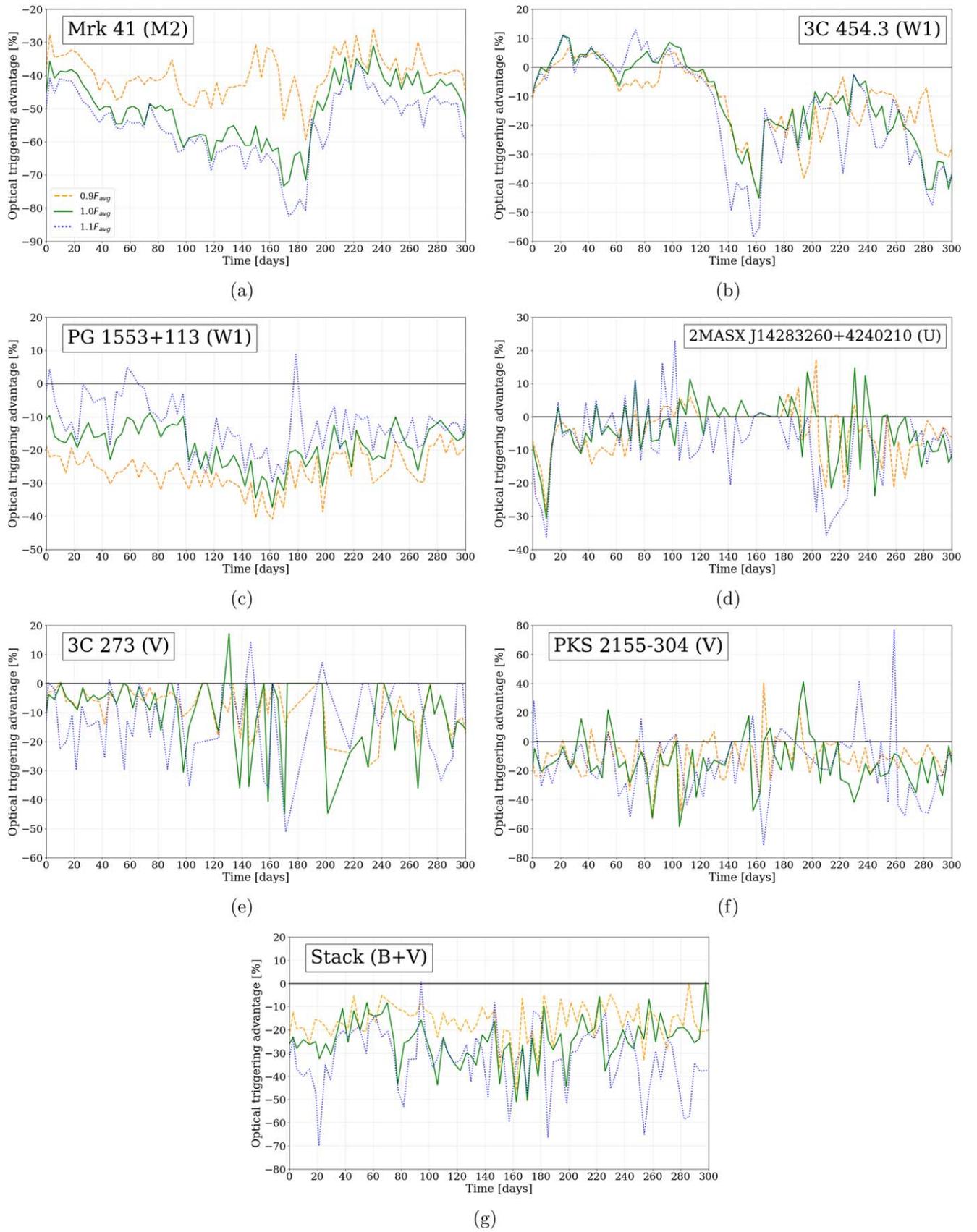

**Figure 8.** The advantage (or disadvantage) of using a UV/optical trigger over an X-ray trigger of the same average flux percentage is illustrated for the top six sources, where the UV/optical-triggered flux is divided by the X-ray-triggered flux to yield the advantage as a percentage.

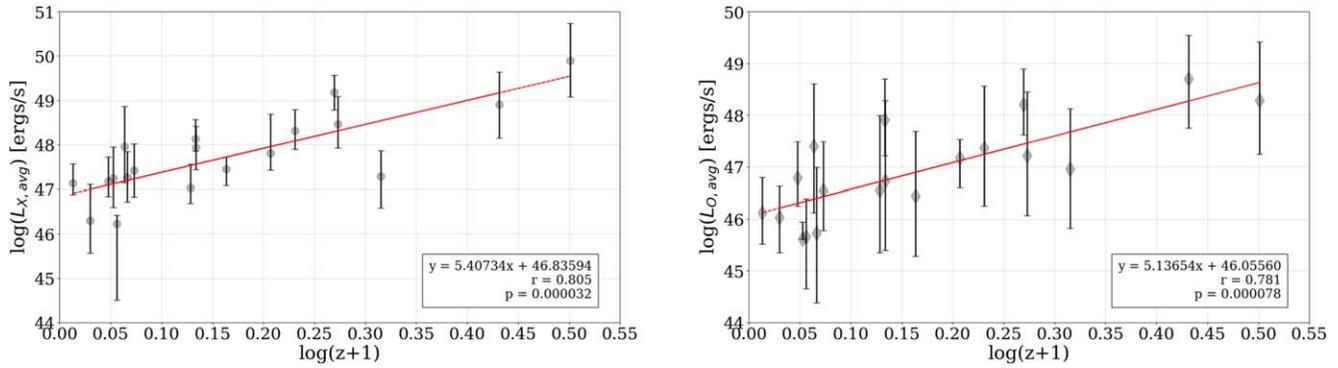

**Figure 9.** Log–log relationship between $1 + z$ and average luminosity (X-ray: left, UV/optical: right). The redshift ranges from 0.031 (Mrk 421) to 2.2 (S5 0836 +71). The Pearson $R$ and probability values of the relationship indicate a strong power-law correlation between the average X-ray luminosity and redshift. There is a moderately strong power-law correlation between the average UV/optical luminosity and redshift. The complete figure set (7 images) is available in the online journal.

(The complete figure set (7 images) is available.)

interferometry studies of the radio jet (Lister et al. 2013), while the weaker (and nearby) source Mrk 421 has has an apparent $v/c < 1$. A more detailed analysis of this line of inquiry would be worthwhile, but is beyond the scope of this paper.

We searched for a relationship between the UV/optical and X-ray fluxes and variabilities in Figure 7. There appears to be a weak positive trend between the X-ray luminosity and UV/optical variability of the 19 sources, and this correlation is less-supported between the UV/optical luminosity and X-ray variability. The latter relationship and errors corresponding to these quantities may arise from the observed variabilities of the sources at X-ray wavelengths (which differ more than the UV/optical variability) and the large range of luminosities exhibited by each source, some of which "flicker" and are difficult to predict following their turnover points (Table 5).

In the structure function analysis, one often finds that a turnover time of an object, calculated in Section 5, approximates the sampling interval in which the structure function of an object flattens. The X-ray and UV/optical turnover times that we compared to average luminosity (in their respective wavelengths) are in the intrinsic frame.

It would be beneficial to compare the turnover points of more objects in the blazar class of AGNs, though a weakly supported correlation may exist in the X-ray band (Figure 7). Figure 7 suggests that there is no correlation between the X-ray and UV/optical turnover points of these sources. Structure function calculations of other flat-spectrum radio sources could also aid our analysis and allow one to characterize AGNs by their variable behavior.

### 7.1. Selectively Observed versus Continuously Monitored Objects

There are three ways an object is observed by Swift: objects scheduled in programs without trigger conditions, calibration sources (sources that are regularly observed because of their brightness), and observations that are triggered by bright outbursts, often targets of opportunity (TOOs). For the first category, there is no bias by brightness; this is also true for the calibration sources. However, a brightness trigger or TOO can provide a biased view of the variability and skew the triggering results, but this information is not retained for such observation. The most extreme outbursts that cause TOOs will produce outliers, which we exclude.

Most sources observed frequently were not TOO objects, according to information provided to us by the Science Operation Center. However, in objects such as Mrk 421, 3C 454.3, PKS 2005-489, and 1RXS J122121.7+301041, up to a third of their observations appear to occur at irregular intervals ranging from one to six or more months, with some exposures occurring more than once a day or for several consecutive days, indicating that they were observed as a TOO for a fraction of their exposures or were not observed due to their brightness.

### 7.2. Mission Planning Implications

Our investigation has a bearing on mission planning for the scientific missions of Arcus, Athena, or Lynx. A goal is to avoid observing targets when they are significantly fainter than normal, which would compromise the signal-to-noise ratio of the observation. The time between scheduling an observation (based on its brightness at the time) and its execution can be months, so we investigated flux variation over time periods of 30 and 100 days.

We utilized the square root of the structure function to quantify an object's variability at a given sampling interval ($\Delta t$). The most variable sources at 30 days (in X-rays) are 1RXS J022716.6+020154 and Mrk 421 (as measured by variabilities exceeding 100% of the previous flux value), followed by 1RXS J122121.7+301041 and 1RXS J150759.8 +041511 (with variabilities exceeding 0.8 normalized flux values). In the UV/optical wavelengths, 3C 454.3 is the most variable object at 30 days and has a variability of about 0.51 normalized flux units. The objects with the next highest variabilities have values that are approximately half of the variability of 3C 454.3 for the same sampling interval, indicating that the UV/optical variability of 3C 454.3 is unusually high compared to other AGNs at 30 days, though these variabilities are more similar across objects in the X-ray band.

The X-ray variability of the targets at 100 days is on average 30% greater than their variability at 30 days (190% greater for UV/optical), and the X-ray flux still appears to vary to a greater extent than the UV/optical. The objects with the highest X-ray and UV/optical variabilities at 100 days (PG 1437+389, Mrk 421, PKS 2155-304, 1RXS J150759.8 +041511, and 3C 273) also exhibit the greatest ranges in flux, especially in the X-ray band, making their flux more difficult to predict. The stack behaves like a typical source at 30 and 100 days, with X-ray and UV/optical variabilities comparable to those of other objects. For both sampling intervals, the X-ray variability is four to five times larger than the UV/optical variability, and the variability at 100 days is 50% and 26% greater than the variability at 30 days for the X-ray and UV/optical bands, respectively.

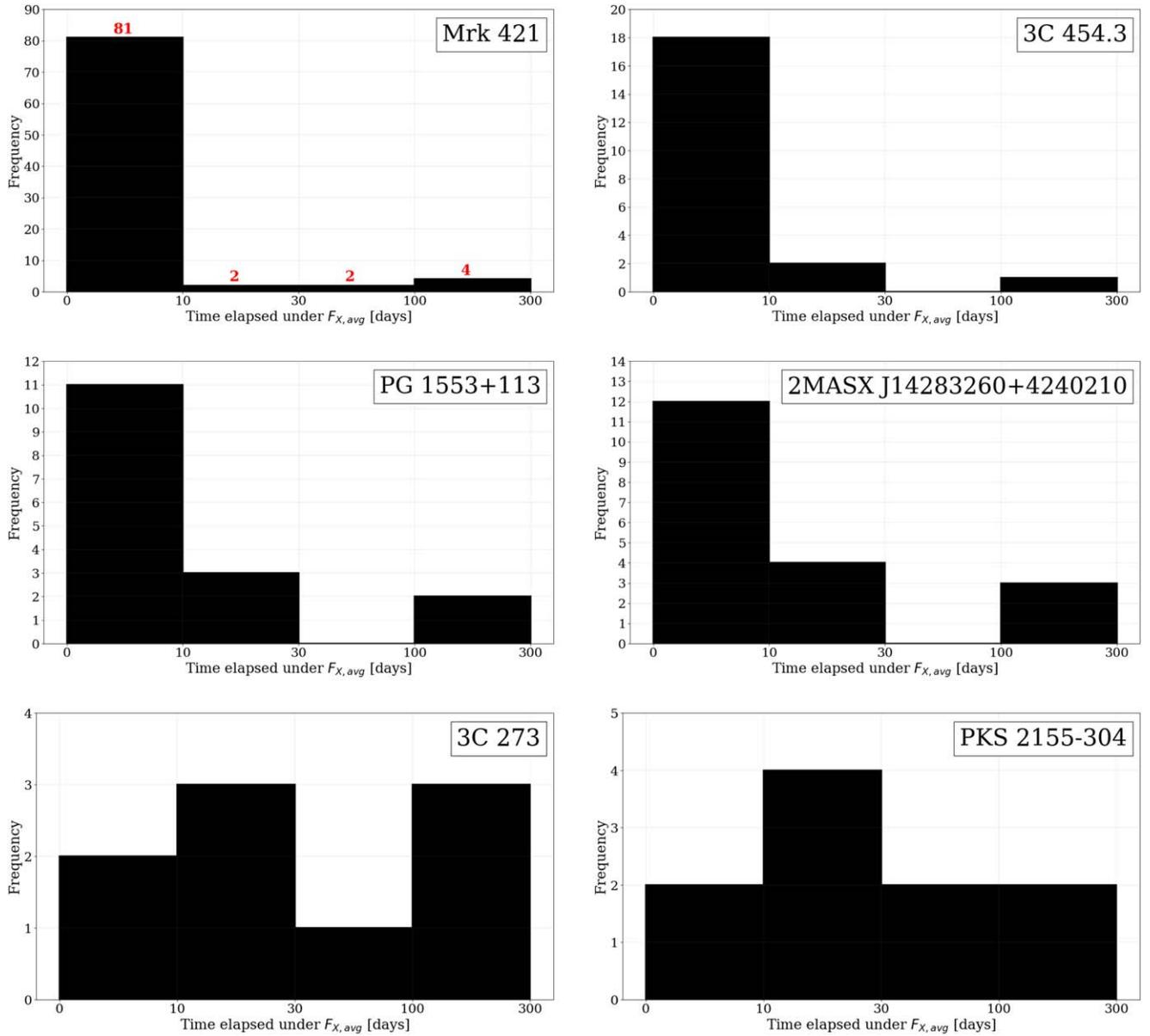

**Figure 10.** Distribution of "low-flux" durations for the top six sources (including outliers). We defined bins with edges of 0, 10, 30, 100, 300, and 1000 days to group occurrences of the flux falling below $F_{X,avg}$ by the length of time until the flux reaches $F_{X,avg}$ again. It appears that a majority of low-flux durations only last from 0 to 10 days, which would not likely affect an observing run of one to three months. Although 3C 273 and PKS 2155-304 feature several low-flux durations that last from 10–30 or 30–100 days, one must take into account the possibility of observing interruptions, which result in a lack of observations for periods of time up to six months.

One wants to avoid observing an object for a period of time in which its flux is lower than its triggering threshold. We investigate the length of time for which the flux of the top six objects remains below the $1.0F_{X,avg}$ triggering threshold in Figure 10.

We utilized X-ray and UV/optical triggering to improve the likelihood of observing an object at an adequately high flux (generally meeting or exceeding the range $0.9F_{X,avg}$–$1.1F_{X,avg}$). Objects with over 100 observations exhibit a more evident brightness advantage when the trigger's percentage of the average flux is increased. Compared to no trigger, fluxes from these objects that are triggered by a value of $0.9F_{X,avg}$ are consistently brighter from 0 to 30 days and remain at an advantage from 30 to 100 days, though this ranges from approximately 5% up to 80% for the six most documented objects. Increasing the triggering value to $1.0F_{X,avg}$ and $1.1F_{X,avg}$ leads to fluxes that are usually around 5%–10% brighter than the flux yielded by the $0.9F_{X,avg}$ trigger, but this comes at the cost of fewer successful triggers.

Generally, it would not be ideal to optically trigger results because they yield lower X-ray fluxes than those that are triggered by X-rays. Out of 19 objects and the stack, one (3C 454.3) suggests that optically triggered X-ray fluxes are brighter than those triggered by X-rays for at least half of the timescale from 0 to 100 days (discussed below); four sources

(2MASX J14283260+4240210, 3C 273, PKS 2155-304, and 1RXS J122121.7+301041) illustrate a temporary or short-lived (25–50 days) advantage to UV/optical triggering, and the rest of the objects (79%, including the stack and its individual objects) do not suggest an advantage. On average, the $0.9F_{\rm O,avg}$ trigger produces fluxes that are 11.7% less luminous than those triggered by $0.9F_{\rm X,avg}$, and these disadvantages of UV/optical triggering for $1.0F_{\rm O,avg}$ and $1.1F_{\rm O,avg}$ are 13.0% and 14.7%, respectively.

3C 454.3 is the only object that illustrates strong advantages of UV/optical triggering. However, as outlined in Section 7.1, there exists the possibility that selective observation (TOO) yields fluxes that are higher than the typical day-to-day behavior of the source, skewing our triggering results higher (even in the absence of outliers). Thus, to eliminate the effect of prompted observations on our triggering analysis, one would need to utilize a data set composed of entirely regularly monitored observations.

Some objects (2MASX J14283260+4240210, 3C 273 in the $B$, M2, W1, W2 bands, PKS 2005-489, and S8 0836+71 in the M2 band) present optically triggered fluxes for a significant time frame (over 30 out of 100 days) that are up to 20% less bright than the X-ray-triggered flux of the same triggering threshold, but still produce brighter flux than utilizing no trigger at all. As there is not a large difference between the triggered fluxes, and UV/optical triggering can be realized on the ground, it may be a viable option to optically trigger X-ray observations to mitigate costs or save time for these targets.

### 7.3. Further Steps

Our 19 sources are primarily blazars, which would be the main targets for extragalactic X-ray absorption line studies. Our understanding of variability behavior will be expanded by extending our X-ray and UV/optical variability analysis and triggering procedure to other classes of AGNs. Geometric factors such as the inclination of the quasar's accretion disk, the direction of the jet, and the size or the mass of the central black hole may change the patterns and correlations seen between the X-ray and UV/optical flux.

One can also trigger observations with other wavelengths of light if it is more cost-effective or easily realized, though UV/optical triggering is usually preferred when observing on the ground. Current databases of AGNs typically have a reasonably large quantity of observations at such wavelengths for well-documented objects, so new, monitored data may be required for a more thorough analysis.

We thank the individuals who offered guidance and insight, including Philip Hughes, Jon Miller, Mateusz Ruszkowski, and an anonymous referee. We are grateful for support from NASA through the Astrophysics Data Analysis Program, awards NNX15AM93G and 80NSSC19K1013. This research has made use of the NASA/IPAC Extragalactic Database (NED), which is operated by the Jet Propulsion Laboratory, California Institute of Technology, under contract with NASA. This research has made use of data and software provided by the High Energy Astrophysics Science Archive Research Center (HEASARC), which is a service of the Astrophysics Science Division at NASA/GSFC and the High Energy Astrophysics Division of the Smithsonian Astrophysical Observatory.

### ORCID iDs

Kaitlyn E. Moo ● https://orcid.org/0000-0003-2753-1461
Joel N. Bregman ● https://orcid.org/0000-0001-6276-9526
Mark T. Reynolds ● https://orcid.org/0000-0003-1621-9392